\documentclass[aps,pra,reprint,superscriptaddress]{revtex4-2}
\usepackage{color}
\usepackage{amsmath}
\usepackage{amssymb}
\usepackage{graphicx}
\usepackage{hyperref}
\usepackage{physics}
\usepackage{times}

\hypersetup{
    colorlinks=true,linkcolor=blue,citecolor=blue,
    filecolor=blue,urlcolor=blue,breaklinks=true
}

\RequirePackage{color}

\newcommand{\orcid}[1]{\href{https://orcid.org/#1}
{\includegraphics[width=7pt]{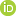}}}

\newtheorem{theorem}{Theorem}

\begin{document}

\title{Self-adjoint extension approach for singular Hamiltonians in
  (2+1) dimensions}

\author{Vinicius Salem\orcid{0000-0002-1768-8783}}
\email{salemfisica@gmail.com}
\affiliation{ICFO - Institut de Ci\`encies Fot\`oniques, Mediterranean Technology Park, 08860 Castelldefels (Barcelona), Spain
}
\affiliation{
  Programa de P\'os-Gradua\c{c}\~ao em Ci\^encias/F\'{i}sica,
  Universidade Estadual de Ponta Grossa,
  84030-900 Ponta Grossa, PR, Brasil
}

\author{Ramon F. Costa}
\email{ramon.fcosta@gmail.com}
\affiliation{
  Programa de P\'os-Gradua\c{c}\~ao em Ci\^encias/F\'{i}sica,
  Universidade Estadual de Ponta Grossa,
  84030-900 Ponta Grossa, PR, Brasil
}

\author{Edilberto O. Silva\orcid{0000-0002-0297-5747}}
\email{edilbertoo@gmail.com}
\affiliation{
  Departamento de F\'{i}sica,
  Universidade Federal do Maranh\~{a}o,
  65085-580 S\~{a}o Lu\'{i}s, MA, Brasil
}

\author{Fabiano M. Andrade\orcid{0000-0001-5383-6168}}
\email{fmandrade@uepg.br}
\affiliation{
  Departamento de Matem\'{a}tica e Estat\'{i}stica,
  Universidade Estadual de Ponta Grossa,
  84030-900 Ponta Grossa, PR, Brasil
}

\date{\today}

\begin{abstract}
In this work, we review two methods used to approach singular
Hamiltonians in (2+1) dimensions.
Both methods are based on the self-adjoint extension approach.
It is very common to find singular Hamiltonians in quantum mechanics,
especially in quantum systems in the presence of topological defects,
which are usually modelled by point interactions.
In general, it is possible to apply some kind of regularization procedure, as the vanishing of the wave function at the location of the
singularity, ensuring that the wave function is square-integrable and
then can be associated with a physical state.
However, a study based on the self-adjoint extension approach can
lead to more general boundary conditions that still gives acceptable
physical states.
We exemplify the methods by exploring the bound and scattering scenarios
of a spin 1/2 charged particle with an anomalous magnetic moment in the
Aharonov-Bohm potential in the conical space.\\
\newline
\href{https://doi.org/10.3389/fphy.2019.00175}
{DOI: 10.3389/fphy.2019.00175}
\end{abstract}

\maketitle

\section{Introduction}
\label{sec:introduction}

Singular and pathological Hamiltonians are quite common in quantum
mechanics and already have a long history \cite{PR.80.797.1950}.
Probably, the first work to deal with $\delta$-like singularities was in
the Kronig-Penny model \cite{PRSLA.130.499.1931}
for the description of the band energy in solid-state physics.
Since then, point interactions have been of great interest in various
branches of physics for their relevance as solvable models
\cite{Book.2004.Albeverio}.
For instance, in the famous Aharonov-Bohm (AB) effect
\cite{PR.115.485.1959} of spin-$1/2$ particles \cite{PRD.40.1346.1989,PRL.64.503.1990,IJMPA.6.3119.1991}
a two-dimensional $\delta$ function appears as the mathematical
description of the Zeeman interaction between the spin and the magnetic
flux tube \cite{PRD.48.5935.1993,AoP.251.45.1996}.
The presence of this $\delta$ function cannot be discarded when the
electron spin is taken into account and it leads to changes in the
scattering amplitude and cross-section
\cite{PRL.64.503.1990}.
This question can also be understood in connection with the quantum
mechanics of a particle in a $\delta$ function potential in one
dimension.
When we wish to solve the problem for bound states, it is well-known
that such a function guarantees at least one bound state
\cite{JAMOP.2011.4.2011,AJP.2002.70.67}, and this property is maintained
when studying the quantum mechanics of other physical systems in the
presence of external magnetic fields.
The inclusion of the spin element in the approach of the AB problem
allows us to establish an exact equivalence with another well-known effect
in the literature, namely the Aharonov-Casher (AC) effect
\cite{PRL.1984.53.319}.
In the AC effect, a spin-1/2 neutral particle with a magnetic moment is
placed in an electric field generated by an infinitely long,
an infinitesimally thin line of charge.
The interaction term involving the particle spin with the electric field
in the AC Hamiltonian is proportional to the $\delta$ function.
Some works in the literature state that point interaction does not
affect the scattering cross-section \cite{JPA.2010.43.354008}.
However, as in the spin-1/2 particle AB problem, the solution of the
equation of motion via the self-adjoint extension in the spin-1/2
neutral particle AC problem reveals that the presence of the $\delta$
function changes the scattering phase shift and consequently the
$S$-matrix \cite{AHEP.2017.2017.7,EPJC.2013.73.2402}.

The study of physical systems with singular Hamiltonians appears in
various contexts of physics.
In Ref. \cite{IJGMMP.2018.15.1850135}, the discrete spectrum of a massive particle trapped in an infinitely long cylinder with two attractive
delta-interactions in the cosmic string spacetime is studied.
The authors showed that the physical effects due to the cosmic string
background are similar to those of the AB effect in quantum mechanics.
This is verified when the cosmic string determines a deviation
on the trajectory of a particle, despite the locally flat character of
the manifold.
In Ref. \cite{PRD.2017.95.045004}, the one-dimensional spinless Salpeter
Hamiltonian with finitely many Dirac delta potentials was solved using
the heat kernel techniques and self-adjoint extension method.
As in the case involving a single $\delta$ potential, the model requires
a renormalization to be made.
They investigated the problem in the context of bound states and showed
that the ground state energy is bounded from below.
Besides, they also showed that there exists a unique self-adjoint
operator associated with the resolvent formula and obtained an explicit
wave function formula for $N$ centres.
The approach using this model to the scattering problem was addressed in
Ref. \cite{EPJ.2018.39.035403}.
Such a model is a  generalization of the work in
Ref. \cite{PRD.2014.89.125023}, where the Schrödinger equation for a
relativistic point particle in an external one-dimensional
$\delta$-function potential was studied using dimensional
regularization.

The physical regularization used in these models is consistent with the
self-adjoint extension theory and the idea can also be used to study
other versions of the Kronig-Penney model in condensed matter physics.
Different forms of Kronig-Penney-type Hamiltonians can be found in
the literature \cite{IEOT.1998.31.436,JPA.1995.28.2313}.
To approach singular Hamiltonian, it is more convenient to apply
von Neumann's theory of self-adjoint extensions
\cite{Book.2004.Albeverio,Book.1975.Reed.II,Book.1993.Akhiezer}.
In general, if we ignore the singularity, the resulting Hamiltonian is
self-adjoint and positive definite \cite{PRA.66.032118.2002}, its
spectrum is  $\mathbb{R}^{+}$ and there are no bound states.
The situation changes if we consider the delta function because the
singularity is physically equivalent to an extraction of a single point
from the plane $\mathbb{R}^2$, which leads to the loss of the
self-adjointeness of the Hamiltonian.
This has important consequences in the spectrum of the system \cite{Book.2008.Oliveira}.
However, the self-adjointness is necessary to have a unitary time
evolution.
So, we must guarantee that the Hamiltonian is self-adjoint, which here
is done employing the self-adjoint extension of symmetric operators.
With this approach, a new family of self-adjoint operators labelled by a
real parameter is obtained.

The situation discussed above occurs, for instance, in the AB scattering
of a spin-1/2 particle, where it is well-known that for all real values
of the self-adjoint extension parameter, there is an additional
scattering amplitude \cite{PRL.64.503.1990}, which results from the
interaction between the spin and the magnetic flux tube
\cite{PRD.85.041701.2012}.
Moreover, there is one bound state solution with negative energy when
this parameter is less than zero.
This situation can be considered quite strange, however, it can be
mathematically proved the existence of this negative eigenvalue
\cite{JRAM.380.87.1987,Book.2004.Albeverio,PRD.40.1346.1989,
CMP.124.229.1989,Book.1995.Jackiw,PLB.267.91.1991,
PLB.333.238.1994,JPA.26.7637.1993,PRD.50.7715.1994,
JMP.36.5453.1995,AoP.323.3150.2008,AoP.325.2529.2010}.
It is interesting to comment that in Ref. \cite{Book.1995.Jackiw}, an
equivalence between the renormalization and the self-adjoint extension
is discussed.

In this paper, we review some elements of the self-adjoint
extension theory which are necessary to address singular
Hamiltonians in relativistic and nonrelativistic quantum theory.
As an application, we consider the model of a spin-1/2 particle with
an anomalous magnetic moment in an AB potential in the cosmic string spacetime.
As already mentioned above, in this model, a $\delta$ function potential
arises in the equation of motion \cite{PR.115.485.1959}.
We derive the Dirac equation for this model and solve it for the
scattering and bound states on the nonrelativistic limit using the
self-adjoint extension method.
The main goal is to study the physical implications of both the cosmic
string background and singularity on the properties of the system.
Our application example is motivated by the importance of studying
cosmic strings \cite{Book.2000.Vilenkin}, which has been the usual
framework for investigating the effects of localized curvature in physical
systems.
There is a significant number of articles in the literature that study
the influence of topology on physical systems using the cosmic string as
a background.

Recently, a detailed study to study geometric phase for an open system
of a two-level atom interacting with a massless scalar field in the
background spacetime of the cosmic string spacetime with torsion was
proposed in Ref. \cite{EPL.2019.126.50005}.
The authors showed that the geometric phase depends not only on the
inherent properties of the atom, but also on the topological properties
of background spacetime.
For this model, it was found that the correction to the geometric phase
of the present system derives from a composite effect, which contains
the cosmic string and screw dislocation associated with the curvature
and torsion, respectively.
The authors also showed that the phase depends on the initial state of
this atom and, in particular, there is no geometric phase acquired for
the atom if the initial state is prepared in the excited state.
Another physical model of current interest that has several studies in
cosmic string spacetime is the Dirac oscillator \cite{JPA.1989.22.817}.
It is known that the Dirac oscillator is a kind of tensor coupling with
a linear potential which leads to the simple harmonic oscillator with a
strong spin-orbit coupling problem in the nonrelativistic limit.
The Dirac oscillator is an exactly soluble model and can be an excellent
example in the context of many-particle models in relativistic and
nonrelativistic quantum mechanics \cite{AIPCP.2011.1334.249}.
In Ref. \cite{EPJC.2019.79.311}, it was studied the relativistic quantum
dynamics of a Dirac oscillator subject to a linear interaction for
spin-1/2 particles in a cosmic string spacetime.
The authors showed in this model that the geometric and topological
properties of these spacetimes lead to shifts in the energy spectrum and
the wave-function.
In Ref. \cite{EPJC.2014.74.3187}, the self-adjoint extension method was
used to study the effects of spin on the dynamics of a two-dimensional
Dirac oscillator in the magnetic cosmic string background.
For other important studies in the cosmic string spacetime, the reader
may refer to the Refs.
\cite{PRL.1989.62.1071,PRD.2010.82.084025,PLA.2007.361.13,PRD.1987.35.3779}
and in the context of nonrelativistic quantum dynamics of a quantum
particle constrained to move on a curved surface using da Costa's
approach \cite{PRA.23.1982.1981} to the Refs.
\cite{JMP.53.122106.2012,AoP.362.739.2015,arXiv:1608.00173}.

The rest of this work is organized as follows.
In Sec. \ref{sec:sae} the theory of the self-adjoint extensions is
presented and two different methods, both based on the self-adjoint
extension, are discussed.
In Sec. \ref{sec:dirac_eq} the Dirac equation that describes the motion
of a spin--1/2 charged particle with an anomalous magnetic moment in the
curved space is developed.
The methods presented in the previous section are then applied to this
system and the scattering and bound states scenarios are discussed.
The scattering matrix and the expression for the bound state energy is
presented.
Finally, in the Sec. \ref{sec:conclusion} we present our conclusions.

\section{The self-adjoint extension approach}
\label{sec:sae}

In this section, we review some important concepts and results from the
von-Neumann-Krein theory of self-adjoint extensions.
Let $A$ and $B$ two operators.
If the domain of $A$ contains the domain of $B$, i.e., $\mathcal{D}(A)
\supseteq \mathcal{D}(B)$, and in the domain of $B$ the operators are
equals, then we say that $A$ is an extension of $B$.
The domain of an operator $A$ is called dense if for each vector $\psi$
in this domain, there is a sequence  $\psi_n$ in such a way that
$\psi_n\to\psi$.
If an operator $A$ has a dense domain, the domain of its adjoint
$A^{\dagger}$, is the set of all vectors $\psi$ for which there is a
vector $A^{\dagger}\psi$ that satisfies
\begin{equation}
  \label{eq:adjoint}
(\phi, A^\dagger \psi) = (A \phi, \psi),
\end{equation}
for all vectors $\phi \in \mathcal{D}(A)$.
Equation \eqref{eq:adjoint} defines  $A^{\dagger}\psi$.
On the other hand, an operator with dense domain $A$ is symmetric if
\begin{equation}
(\phi, A \psi) = (A \phi, \psi),
\end{equation}
for every $\phi$ and $\psi$ in its domain.
In this case $A^{\dagger}\psi$ is defined as  $A^\dagger \psi = A \psi$ for
all $\psi \in \mathcal{D}(A)$, and $A^\dagger$ is said to be an
extension of $A$.
If $A^\dagger = A$, then $A$ is called self-adjoint or Hermitian.
It is interesting to comment that in physics it is common to assume
that Hermitian is the same as self-adjointness.
However, they are different notions in mathematics literature and only
the word Hermitian could be used for symmetric.

An important point here is that a symmetric operator can fail to be a
self-adjoint operator.
For $A$ to be a self-adjoint operator it has to be symmetric,
$A=A^{\dagger}$, and the domains of the operator and its adjoint have to
be equal as well, $\mathcal{D}(A) = \mathcal{D}(A^{\dagger})$.
So, in the same way as a function needs a rule, a domain and a codomain to
be defined, an operator needs not only its action but also its
domain (Hilbert space) to be completely defined.
Several traditional textbooks on quantum mechanics
\cite{Book.2011.Sakurai,Book.1977.Cohen-Tannoudji,Book.1994.Shankar,
  Book.2003.Gasiorowicz}
do not mention the problems that could arise by the incorrect or
incomplete definition of the operators.
An exception being the textbook of the author Ballentine
\cite{Book.1998.BallentineLeslie}.
The mathematical framework of quantum mechanics is that of linear
operators in Hilbert spaces and the problems and paradoxes that could
arise come from the use of simplified rules described in many
textbooks.
As an example of this is the use of the theory if bounded operators to
deal with unbounded operators \cite{AJP.69.322.2001}.

\subsection{The Weyl-Von Neumann's theorem}
\label{subsec:von_neumann}

Following the concept of self-adjoint extension, the question we want to answer is how many extensions, if any, are admitted by an operator.
The answer to this question lies in the concept of deficiency index of
an operator.
Let $A$ be a symmetric operator with domain $\mathcal{D}(A)$ and the
corresponding adjoint operator $A^{\dagger}$ with domain
$\mathcal{D}(A^\dagger)$.
The deficiency subspaces $\mathcal{N}_\pm$ are defined by
\cite{AJP.69.322.2001}
\begin{equation}
  \label{eq:Nmm}
  \mathcal{N}_{\pm} = \left\lbrace \psi_{\pm} \in \mathcal{D}(A^\dagger),
    \quad
    A^\dagger \psi_{\pm} = z_{\pm}\psi_{\pm},
    \quad
    \Im(z_{\pm}) \gtrless 0 \right\rbrace,
\end{equation}
with dimensions $\dim{\{\mathcal{N}_{\pm}\}}=n_{\pm}$.
The pair of  nonnegative integers $(n_{+},n_{-})$ are called
deficiency indices of $A$.
The exact value of $z_{\pm}$ is not important as long as $z_{+}$
($z_{-}$) belongs to the upper (lower) half complex plane.
For simplicity, it is chosen as $z_{\pm}=\pm i z_{0}$, with
$z_0$ an arbitrary positive real number, used for dimensional reasons.
In this manner, to access the deficiency indices, all we have to do is
to solve the eigenvalue equation
\begin{equation}
  \label{eq:defspaces}
A^\dagger \psi_{\pm} = \pm i z_{0} \psi_{\pm},
\end{equation}
and then count the number of linearly independent solutions that
belong to the domain of the adjoint operator in the Hilbert space in
question, i.e., those that are square integrable.
\begin{theorem}[Weyl and Von Neumann \cite{AJP.69.322.2001}]
  \label{th:WVN}
  Consider an operator $A$ with deficiency index $(n_{+},n_{-})$:
  \begin{enumerate}
  \item If $n_+ = n_-$, $A$ is essentially self-adjoint;
  \item If $n_+ = n_- = n \geq 1$, $A$ posses an infinity number of
    self-adjoint extensions parametrized by a unitary matrix
    $U:\mathcal{N}_{+}\to \mathcal{N}_{-}$ of dimension
    $n$ with $n^2$ real parameters;
  \item If $n_+ \neq n_-$,  $A$ does not admit a self-adjoint extension.
  \end{enumerate}
\end{theorem}
Therefore, the domain of $A^{\dagger}$ is
\begin{equation}
  \mathcal{D}(A^{\dagger}) = \mathcal{D}(A)
  \oplus \mathcal{N}_{+}\oplus \mathcal{N}_{-}.
\end{equation}
So, it is important to note that even for Hermitian operators,
$A=A^{\dagger}$, its domains might be different.
In this manner, the self-adjoint extension essentially consists of
extending the domain of $A$ using the deficiency subspaces
$\mathcal{N}_{\pm}$ to match the domain of $A^{\dagger}$.

Now that we have discussed some general concepts about the self-adjoint
extension approach, we restrict our discussion to the specific case of
singular Hamiltonian operators $H$ in (2+1) dimensions.
In these cases, the singularity is characterized by the presence of a
two-dimensional $\delta$ function localized at the $r=0$.
It is well-known in the literature that these Hamiltonians are not
self-adjoint and admit a one-parameter family of self-adjoint extension
\cite{Book.1975.Reed.II}.
Thus, our main goal is to solve the time-independent Schr\"{o}dinger
equation
\begin{equation}
  H \psi = E \psi,
\end{equation}
with $H$ the Hamiltonian, $\psi$ the wave function and $E$ the energy.
To do so, we shall discuss two methods to characterize the
family of self-adjoint extensions of $H$.
In both methods, the delta function singularity is replaced by a boundary condition at the origin.
In the first one, proposed by Bulla and Gesztesy (BG) in
\cite{JMP.26.2520.1985}, the boundary condition is a mathematical limit
allowing divergent solutions for the Hamiltonian $H$ at isolated points,
provided they remain square-integrable.
In the second one, proposed by Kay and Studer (KS) in
\cite{CMP.139.103.1991}, the boundary condition is a match of the
logarithmic derivatives of the zero-energy solutions for
the regularized Hamiltonian and the solutions for the
Hamiltonian $H$ without the delta function plus a self-adjoint
extension.
As we shall show, the comparison between the results of the two methods
allows us to express the self-adjoint extension parameter (a mathematical parameter that characterizes the self-adjoint extension) in
terms of the physics of the problem.

\subsection{The BG  method}
\label{subsec:bg}

Let us consider the radial singular Schr\"{o}dinger operator in
$L^2((0,\infty))$ given by
\begin{equation}
  \label{eq:hBG}
    h = - \frac{d^2}{dr^2} + \frac{\ell(\ell - 1)}{r^2}
    + \frac{\gamma}{r} + \frac{\beta}{r^{a}} + W,
\end{equation}
with $W \in L^{\infty}((0,\infty))$ real valued and
$1/2 \leq \ell < 3/2$, $\beta,\gamma \in \mathbb{R}$, $0<a<2$.
Bulla and Gesztesy showed that this operator, in the interval
$1/2 \leq\ell < 3/2$, is not self-adjoint having deficiency indices
$(1,1)$.
Thus admitting a one-parameter family of self-adjoint extensions.
The following theorem characterizes all the self-adjoint extension of
$h$.
\begin{theorem}
  \label{th:BG}
  [Bulla and Gesztesy \cite{JMP.26.2520.1985,Book.2004.Albeverio}]
  All the self-adjoint extension $h_{\nu}$ of $h$ can  be characterized
  by
  \begin{equation}
    h_{\nu}= - \frac{d^2}{dr^2} + \frac{\ell(\ell - 1)}{r^2}
    + \frac{\gamma}{r} + \frac{\beta}{r^{a}} + W,
    \end{equation}
with domain
\begin{align}
  \label{eq:eq_dom_ext_bg}
  \mathcal{D}(h_\nu) =
  &
    \big\{g \in L^2 \left(\left( 0, \infty \right)\right)\big \rvert
    g, g' \in AC_{\rm loc} \left(\left( 0, \infty \right)\right);  \\
  & -g'' + \frac{\ell(\ell - 1)}{r^2}g + \frac{\gamma}{r}g +
    \frac{\beta}{r^{a}}g \in L^2 \left(\left( 0, \infty \right)\right)
    \big\}
\end{align}
with $AC_{\rm loc}((a,b))$ denoting the set of locally absolutely
continuous functions on $((a,b))$ and the function $g$ satisfies the boundary
condition
\begin{equation}
  \label{eq:cc_bg}
\nu g_{0, \ell} = g_{1, \ell},
\end{equation}
and
\begin{equation}
  \label{eq:bg_constantes}
  -\infty < \nu \le \infty,  \quad
  \frac{1}{2} \le \ell < \frac{3}{2}, \quad
  \beta, \gamma \in \mathbb{R}, \quad 0 < a < 2.
\end{equation}
The boundary values in \eqref{eq:cc_bg} are defined by
\begin{equation}
  \label{eq:cc_bg_0}
  g_{0, \ell} = \lim_{r \to 0^+} \frac{g(r)}{G^{(0)}_{\ell}(r)},
\end{equation}
and
\begin{equation}
  \label{eq:cc_bg_1}
  g_{1, \ell} = \lim_{r \to 0^+}
  \frac{g(r) - g_{0, \ell} G^{B}_{\ell}(r)}
  {F ^{(0)}_{\ell}(r)}.
\end{equation}
The boundary condition $g_{0,\ell}=0$ (i.e., $\nu=\infty$) represents the
Friedrichs extension of $h$.
\end{theorem}

The functions $F ^{(0)}_{\ell}(r)$ and $G^{(0)}_{\ell}(r)$ are given by
\begin{equation}\label{eq:eq_bg_f}
F ^{(0)}_{\ell}(r) = r^\ell,
\end{equation}
and
\begin{equation}\label{eq:eq_bg_g}
  G ^{(0)}_{\ell}(r) =
  \begin{cases}
    -r^{1/2} \ln (r),
    & \ell = \frac{1}{2}, \\
    \displaystyle \frac{r^{1 - \ell}}{(2 \ell - 1)},
    &  \frac{1}{2} < \ell < \frac{3}{2}.
  \end{cases}
\end{equation}
$G^{B}_{\ell}(r)$ denotes the asymptotic expansion fo $G_\ell(r)$ for
$r \to 0^+$ up to $r^t$, with $t \le 2 \ell - 1$.

\subsection{The KS method}
\label{subsec:kay_studer}

The authors Kay and Studer studied, in the context of self-adjoint
extensions, the boundary conditions for singular Hamiltonians in conical
spaces and fields around cosmic strings \cite{CMP.139.103.1991}.
Among the studied problems, are the AB like problems in two
dimensions.

The KS method starts by considering a regularization procedure for the
point interaction at the origin.
Thus, for the regularized Hamiltonian, where the point interaction is
shifted from the origin by a finite very small radius $r_{0}$, the method is applied in the following manner \cite{AoP.339.510.2013}:
\begin{enumerate}
\item We temporally forget the point interaction at the origin
  substituting the singular Hamiltonian by the corresponding nonsingular
  one;

\item We solve the Eq. \eqref{eq:defspaces} for the deficiency spaces
  of the nonsingular Hamiltonian;

\item The solutions obtained in the previous step are used to complete
  the space of solutions for the nonsingular Hamiltonian;

\item In the last step, a boundary condition matching the logarithmic
  derivatives of the zero-energy solutions for the regularized
  Hamiltonian of step 1 and the general solutions obtained in step 3 is
  employed:
  \begin{equation}
    \label{eq:eq_cc_ks}
  \lim_{r_0 \to 0^+} r_0 \frac{\dot{g}_{0}}{g_0} \biggr \vert_{r = r_0} =
  \lim_{r_0 \to 0^+} r_0 \frac{\dot{g}_\rho}{g_\rho} \biggr \vert_{r = r_0}.
\end{equation}
In the above equation, $g_\rho$ are the solutions obtained in step 3 and
$g_0$ are the zero-energy solutions ($\dot{g}=d g/dr$).
\end{enumerate}

Now that we have discussed the self-adjoint extension approach and the
BG and KS methods, in what follows we exemplify the application of both
methods to the problem of a spin--1/2 charged particle with an anomalous magnetic moment under the influence of an AB field in conical space.

\section{The Dirac equation for the AB system in the conical space}
\label{sec:dirac_eq}

In this section, we shall obtain the Dirac equation to describe the motion
of a spin--$1/2$ charged particle with mass $M$ and anomalous magnetic
moment $\mu _{B}$ interacting with an AB field in the cosmic string
spacetime.
The line element that describes this universe written in
cylindrical coordinates is given by
\begin{equation}
  ds^{2}=dt^{2}-dr^{2}-\alpha ^{2}r^{2}d\varphi ^{2}-dz^{2},
  \label{eq:metric}
\end{equation}
with $-\infty <(t,z)<\infty $, $r\geq 0$ and $0\leq \varphi \leq 2\pi $.
The
parameter $\alpha $ in the metric \eqref{eq:metric} is related to the linear
mass density $\bar{m}$ of the cosmic string through the formula $\alpha =1-4
\bar{m}$ and it stands for two situations:

\begin{itemize}
\item
  It describes the surface of a cone if $0<\alpha <1$.
  This is equivalent to removing a wedge angle of $2\pi (1-\alpha )$ and
  the defect presents a positive curvature.

\item
  It describes the surface of an anticone or the figure of a
  saddle-like surface when $\alpha >1$.
  This situation corresponds to the addition of an excess angle of $2\pi
  (\alpha -1)$ and, in this case, the defect represents a negative
  curvature.
\end{itemize}
In this work, we shall discuss the case of a conical surface, so that $0
< \alpha \leq 1$, with the equality corresponding to the flat space.

The metric in \eqref{eq:metric} can also be read as the Minkowski spacetime
with a conic singularity at $r=0$ \cite{SPD.22.312.1977}.
Because of this characteristic, the only nonzero components of the
curvature tensor is found to be
\begin{equation}
  R_{r,\varphi }^{r,\varphi} =
  \frac{1-\alpha }{4\alpha }\delta _{2}(\mathbf{r}),
  \label{eq:curva}
\end{equation}
where $\delta _{2}(\mathbf{r})$ is the two-dimensional delta function in
flat space.
The conical singularity in the tensor \eqref{eq:curva} reveals that
the curvature is concentrated on the cosmic string axis and in all other
regions it is null.

Since the spacetime is not flat, we must take into account the spin
connection in the Dirac equation.
To implement this, we need to construct a frame which allows us to
obtain the Dirac gamma matrices $\gamma ^{\mu }$ in
the Minkowskian spacetime (defined in terms of the local coordinates) in
terms of global coordinates.
This is done by using the tetrad base
$e_{\,\,\,\mu }^{\left( a\right) }\left( x\right)$, which allows to
contract the matrices $\gamma ^{\mu }$ with the inverse tetrad
$e_{\left( a\right) }^{\mu }\left( x\right) $ through the relation
\begin{equation}
\gamma ^{\mu }\left( x\right) =e_{\left( a\right) }^{\mu }\left( x\right)
\gamma ^{\left( a\right) },  \label{eq:gmatrices}
\end{equation}
satisfying the generalized Clifford algebra
\begin{equation}
\left\{ \gamma ^{\mu }\left( x\right) ,\gamma ^{\nu }\left( x\right)
\right\} =2g^{\mu \nu }\left( x\right) ,
\end{equation}
with
\begin{equation}
g_{\mu \nu }\left( x\right) =e_{\,\,\,\mu }^{\left( a\right) }\left(
x\right) \,e_{\,\,\,\nu }^{\left( b\right) }\left( x\right) \,\eta _{\left(
a\right) \left( b\right) },  \label{2.2}
\end{equation}
being the metric tensor of the spacetime in the presence of the
background topological defect, where
$\eta _{\left( a\right) \left( b\right) }$ is the metric tensor of the
flat space, and $(\mu ,\nu )=(0,1,2,3)$ represent tensor indices while
$(a,b)=(0,1,2,3)$ are tetrad indices.
The tetrad and its inverse satisfy the following properties:
\begin{equation}
e_{\,\,\,\mu }^{\left( a\right) }\,\left( x\right) e_{\,\,\,\left( b\right)
}^{\mu }\left( x\right) =\delta _{\,\,\,\left( b\right) }^{\left( a\right)
}\,\,\,\,\,\,\,e_{\,\,\left( \,a\right) }^{\mu }\left( x\right)
\,e_{\,\,\,\nu }^{\left( a\right) }\left( x\right) =\delta _{\,\,\,\nu
}^{\mu }.
 \label{2.3a}
\end{equation}
The matrices
$\gamma ^{\left( a\right) }=
\left( \gamma ^{\left( 0\right)},\gamma ^{\left( i\right) }\right) $
in Eq. \eqref{eq:gmatrices} are the standard Dirac matrices in Minkowski
spacetime, those representation is
\begin{equation}
\gamma ^{\left( 0\right) }=\left(
\begin{array}{rr}
I & 0 \\
0 & \mathbb{-}I
\end{array}
\right) ,~~~\gamma ^{\left( i\right) }=
\left(
\begin{array}{cc}
0 & \sigma ^{i} \\
-\sigma ^{i} & 0
\end{array}
\right) ,\quad (i=1,2,3),  \label{eq:standard}
\end{equation}
where $\sigma ^{i}=\left( \sigma ^{1},\sigma ^{2},\sigma ^{3}\right) $ are
the standard Pauli matrices and $I$ is the $2\times 2$ identity matrix.

To write the generalized Dirac equation in the cosmic string background, we
have to take into account the minimal and nonminimal couplings of the
spinor to the electromagnetic field embedded in a classical gravitational
field.
The Dirac equation then reads
\begin{multline}
\Big[ i\gamma ^{\mu }\left( x\right) \left( \partial _{\mu }+\Gamma _{\mu
}\left( x\right) \right) -e\gamma ^{\mu }\left( x\right) A_{\mu }\left(
x\right) \\
-\frac{a_{e}\mu _{B}}{2}\sigma ^{\mu \nu }\left( x\right) F_{\mu
\nu }\left( x\right) -M\Big] \Psi \left( x\right) =0,  \label{eq:diracg}
\end{multline}
where $e$ is the electric charge,
\begin{equation}
a_{e}=\frac{g_{e}-2}{2}=0.00115965218091,  \label{eq:afactor}
\end{equation}
is the anomalous magnetic moment defined, with $g_{e}$ being the
electron's
$g$-factor \cite{RMP.88.035009.2016},
\begin{equation}
A_{\mu }\left( x\right) =\left( A_{0},-\mathbf{A}\right),
\end{equation}
is the 4-potential of the external electromagnetic field, with
$\mathbf{A}$ being the vector potential and $A_{0}$ the scalar
potential,
\begin{equation}
F_{\mu \nu }=\partial _{\mu }A_{\nu }-\partial _{\nu }A_{\mu },
\end{equation}
is the electromagnetic field tensor whose components are given by
\begin{equation}
\left( F_{0i},F_{ij}\right) =\left( E^{i},\varepsilon _{ijk}B^{k}\right), \label{eq:field}
\end{equation}
and the operator
\begin{align}
  \sigma ^{\mu \nu }(x) = {}
  &
    \frac{i}{2}[e_{\left( a\right) }^{\mu}(x) \gamma
    ^{( a) },e_{ (b) }^{\nu}( x) \gamma^{( b) }]\nonumber \\
    = {}
  &
    \frac{i}{2}
    \left[ e_{(a) }^{\mu }
    \gamma ^{( a) }e_{( b) }^{\nu }(x)
    \gamma ^{(b) }-e_{( b) }^{\nu }(x)
    \gamma ^{(b) }e_{( a) }^{\mu }(x)
    \gamma ^{(a) }\right],
\end{align}
those components are given by
\begin{align}
\sigma ^{0i} = {} & i\alpha ^{j}=i\left(
\begin{array}{cc}
0 & \sigma ^{i} \\
\sigma ^{i} & 0
\end{array}
\right) ,  \label{eq:sigmai} \\
\sigma ^{ij} = {} &-\epsilon _{ijk}\Sigma ^{k}=-\left(
\begin{array}{cc}
\epsilon _{ijk}\sigma ^{k} & 0 \\
0 & \epsilon _{ijk}\sigma ^{k}
\end{array}
\right) ,
\end{align}
where
\begin{equation}
\Sigma ^{k}=\left(
\begin{array}{cc}
\sigma ^{k} & 0 \\
0 & \sigma ^{k}
\end{array}
\right)
\end{equation}
is the spin operator.
The spinor affine connection in Eq. \eqref{eq:diracg} is
related with the tetrad fields as \cite{APPB.41.1827.2010}
\begin{equation}
\Gamma_{\mu}\left( x\right) =\frac{1}{8}\omega _{\mu \left( a\right)
\left( b\right) }\left( x\right) \left[ \gamma ^{\left( a\right) },\gamma
^{\left( b\right) }\right] ,  \label{eq:conexs}
\end{equation}
where $\omega _{\mu \left( a\right) \left( b\right) }$ is the spin
connection, which can be calculated from the relation
\begin{align}
  \omega _{\mu \left( a\right) \left( b\right) }\left( x\right) = {}
  &
    \eta _{\left(
a\right) \left( c\right) }e_{\nu }^{\left( c\right) }\left( x\right)
    e_{\left( b\right) }^{\tau }\left( x\right) \Gamma _{\tau \mu }^{\nu
    }
    \nonumber \\
  &
    -\eta
_{\left( a\right) \left( c\right) }e_{\nu }^{\left( c\right) }\left(
x\right) \partial _{\mu }e_{\left( b\right) }^{\nu }\left( x\right) ,
\label{eq:conx}
\end{align}
and $\Gamma _{\tau \mu }^{\nu }$ are the Christoffel symbols.

Now, we need of the tetrad fields to write the Dirac equation in curved
space.
For the cosmic string spacetime they are chosen to be
\cite{PRD.78.064012.2008}
\begin{align}
  e_{\,\,\,\mu }^{(a)}= {}
  &
    \left(
\begin{array}{cccc}
1 & 0 & 0 & 0 \\
0 & \cos \varphi & -\alpha r\sin \varphi & 0 \\
0 & \sin \varphi & \alpha r\cos \varphi & 0 \\
0 & 0 & 0 & 1
\end{array}
            \right) ,\nonumber \\
  e_{\,\,\,(a)}^{\mu }= {}
  &
    \left(
\begin{array}{cccc}
1 & 0 & 0 & 0 \\
0 & \cos \varphi & \sin \varphi & 0 \\
0 & -\sin \varphi /\alpha r & \cos \varphi /\alpha r & 0 \\
0 & 0 & 0 & 1
\end{array}
\right) .
 \label{eq:tetrad}
\end{align}
Using \eqref{eq:tetrad}, the matrices $\gamma ^{\mu }\left( x\right) $
in Eq. \eqref{eq:gmatrices} are written more explicitly as
\begin{eqnarray}
\gamma ^{0} &=&\beta \equiv \gamma ^{t},  \label{eq:ga0} \\
\gamma ^{z} &\equiv &\gamma ^{z},  \label{eq:gaz} \\
\gamma ^{1} &\equiv &\gamma ^{r}=\gamma ^{\left( 2\right) }\cos \varphi
+\gamma ^{\left( 2\right) }\sin \varphi ,  \label{eq:gr} \\
\gamma ^{2} &\equiv &\frac{\gamma ^{\varphi }}{\alpha r}=\frac{1}{\alpha r}
\left( -\gamma ^{\left( 1\right) }\sin \varphi +\gamma ^{\left( 2\right)
}\cos \varphi \right) ,  \label{eq:gphi} \\
\gamma ^{3} &\equiv &\gamma ^{z}.
\end{eqnarray}
The matrices \eqref{eq:ga0}-\eqref{eq:gphi} satisfy condition $\nabla _{\mu
}\gamma ^{\mu }=0$, which means that they are covariantly constant.
The
Pauli matrices $\sigma ^{i}$ in Eq. \eqref{eq:sigmai} have the following
representation:
\begin{eqnarray}
\sigma ^{r} &=&\left(
\begin{array}{cc}
0 & e^{-i\varphi } \\
e^{i\varphi } & 0
\end{array}
\right) , \\
\sigma ^{\varphi } &=&\frac{1}{\alpha r}\left(
\begin{array}{cc}
0 & -ie^{-i\varphi } \\
ie^{i\varphi } & 0
\end{array}
\right) .
\end{eqnarray}

Using the basis tetrad \eqref{eq:tetrad}, the affine connection \eqref{eq:conexs}
is found to be \cite{EPJC.2019.79.596}
\begin{equation}
  \boldsymbol{\Gamma }=\left( 0,0,\Gamma _{\varphi },0\right) ,
  \label{eq:connec}
\end{equation}
where
\begin{equation}
\Gamma _{\varphi }=\frac{1}{2}\left( 1-\alpha \right) \gamma _{\left(
1\right) }\gamma _{\left( 2\right) }=-i\frac{\left( 1-\alpha \right) }{2}
\sigma ^{z},  \label{eq:gammaphi}
\end{equation}
arises as the only nonzero component.

For simplicity, let us assume that the particle interacts with the AB
potential, which is generated by a solenoid along the $z$ direction.
Since the motion is translationally invariant along this direction, we
require that $p_{z}=z=0$ and, in Eq. \eqref{eq:field}, we take $E^{i}=0$
for $i=1,2,3$.
Thus, the particle has a purely planar motion.
Equation \eqref{eq:diracg} takes the form
\begin{multline}
  \Big\{ -i\partial _{0}+\mathbf{\alpha }\cdot
    \Big[ \frac{1}{i}(\mathbf{\nabla }_{\alpha }
      +\mathbf{\Gamma })-e\mathbf{A}\Big]
   \\
    -a_{e}\mu_{B}\gamma^{0}\mathbf{\Sigma }\cdot\mathbf{B}
    +\gamma ^{0}M\Big\}\Psi (x)=0.
\label{eq:dn3}
\end{multline}
It is well-known that, in the nonrelativistic limit, the large energy
$M$ is the driving term in Eq. \eqref{eq:dn3}.
So, writing
\begin{equation}
\Psi =e^{-iEt}\left(
\begin{array}{c}
\chi  \\
\Phi
\end{array}
\right) ,
\end{equation}
we obtain the coupled equations system
\begin{align}
\mathbf{\sigma }\cdot \left[ \frac{1}{i}(\mathbf{\nabla }_{\alpha }+\mathbf{
  \Gamma })-e\mathbf{A}\right] \Phi  = {}
  &\left( i\partial _{0}+a_{e}\mu _{B}
\mathbf{\sigma }\cdot \mathbf{B}\right) \chi ,  \label{eq:n1} \\
\mathbf{\sigma }\cdot \left[ \frac{1}{i}(\mathbf{\nabla }_{\alpha }+\mathbf{
  \Gamma })-e\mathbf{A}\right] \chi  = {}
  &\left( i\partial _{0}-a_{e}\mu _{B}
\mathbf{\sigma }\cdot \mathbf{B}+2M\right) \Phi .  \label{eq:n2}
\end{align}
On the right side of Eq. \eqref{eq:n2}, if we assume that $2M\gg \left(
i\partial _{0}-a_{e}\mu _{B}\mathbf{\sigma }\cdot \mathbf{B}\right) $, we
solve it as
\begin{equation}
\Phi =\frac{1}{2M}\mathbf{\sigma }\cdot \left[ \frac{1}{i}(\mathbf{\nabla }+
\mathbf{\Gamma })-e\mathbf{A}\right] \chi .  \label{eq:ap}
\end{equation}
Substituting \eqref{eq:ap} into \eqref{eq:n1}, we get
\begin{multline}
\frac{1}{2M}\mathbf{\sigma }\cdot \left[ \frac{1}{i}(\mathbf{\nabla }
_{\alpha }+\mathbf{\Gamma })-e\mathbf{A}\right] \mathbf{\sigma }\cdot \left[
\frac{1}{i}(\mathbf{\nabla }+\mathbf{\Gamma })-e\mathbf{A}\right] \chi
\\
-a_{e}\mu _{B}\mathbf{\sigma }\cdot \mathbf{B}\chi =i\partial _{0}\chi .
\label{eq:diracn}
\end{multline}
Using the relation for Pauli's matrices
\begin{equation}
\left( \mathbf{\sigma }\cdot \mathbf{a}\right) \left( \mathbf{\sigma }\cdot
\mathbf{b}\right) =\mathbf{a}\cdot \mathbf{b}+i\mathbf{\sigma }\cdot \left(
\mathbf{a}\times \mathbf{b}\right) ,
\end{equation}
where $\mathbf{a}$ and $\mathbf{b}$ are arbitrary vectors,
Eq. \eqref{eq:diracn} becomes
\begin{equation}
\frac{1}{2M}\left[ \frac{1}{i}(\mathbf{\nabla }_{\alpha }+\mathbf{\Gamma })-e
\mathbf{A}\right] ^{2}\chi -\frac{e}{2M}\left( 1+a_{e}\right) \mathbf{\sigma
}\cdot \mathbf{B}\chi =i\partial _{0}\chi.  \label{eq:cst}
\end{equation}
Now we need to define the field configuration.
We consider the magnetic field generated by an infinity long cylindrical
solenoid in the metric \eqref{eq:metric}.
Thus, in the Coulomb gauge, the vector potential reads
\begin{equation}
  e\mathbf{A}=-\frac{e\Phi }{2\pi \alpha r}\boldsymbol{\hat{\varphi}}
  =-\frac{\phi }{\alpha r}\boldsymbol{\hat{\varphi}},\qquad A_{0}
  =0,  \label{eq:fieldA}
\end{equation}
and
\begin{equation}
  e\mathbf{B}
  =-\frac{e\Phi }{2\pi \alpha}\frac{\delta
    (r)}{r}\mathbf{\hat{z}}
  =-\frac{\phi }{\alpha }\frac{\delta (r)}{r}\mathbf{\hat{z}},
  \label{eq:fieldB}
\end{equation}
with $\phi =\Phi /\Phi _{0}$ being the magnetic flux and $\Phi _{0}=2\pi /e$
is the quantum of magnetic flux.
As we can observe, this magnetic field is
singular at the origin.
The presence of this singularity (a point
interaction) in the Hamiltonian, demands that it must be treated by some
kind of regularization or, more appropriately, by using the self-adjoint
extension approach.
We can note that $\chi $ in Eq. \eqref{eq:cst} is an
eigenfunction of $\sigma ^{z}$, with eigenvalues $s=\pm 1$.
In this way, we
can write $\sigma ^{z}\chi =\pm \chi =s\chi $.
We can take the solutions in
the form
\begin{equation}
\chi \left( t,r,\varphi \right) =e^{-iEt}\left(
\begin{array}{c}
\chi_{+}\left( r,\varphi \right)  \\
\chi _{-}\left( r,\varphi \right)
\end{array}
\right) =e^{-iEt}\chi_{s}\left( r,\varphi \right) .
 \label{eq:ans}
\end{equation}
Substituting \eqref{eq:gammaphi}, \eqref{eq:fieldA}, \eqref{eq:fieldB} and \eqref{eq:ans}
) in Eq. \eqref{eq:cst}, we obtain
\begin{multline}
\frac{1}{2M}\left[ \frac{1}{i}\mathbf{\nabla }_{\alpha }-\frac{(1-\alpha )}{
2\alpha r}s\boldsymbol{\hat{\varphi}}+\frac{\phi }{\alpha r}\boldsymbol{\hat{
  \varphi}}\right] ^{2}\chi_{s}
\\
+\frac{1}{2M}\frac{g_{e}s\phi }{2\alpha }\frac{
\delta (r)}{r}\chi_{s}\left( r,\varphi \right) =E\chi_{s}\left( r,\varphi
\right) .
\end{multline}
Therefore, the eigenvalues equation associated with Eq. \eqref{eq:diracg} is ($k^{2}=2ME$)
\begin{equation}
H\chi _{s}=k^{2}\chi_{s},
\end{equation}
with
\begin{equation}
  H=
  \left[ -i\boldsymbol{\nabla }_{\alpha }
    -\frac{\left( 1-\alpha \right) }
    {2\alpha r}s\boldsymbol{\hat{\varphi}}
    +\frac{\phi }{\alpha r}
    \boldsymbol{\hat{\varphi}}\right] ^{2}
  +\frac{g_{e}s\phi }{2\alpha }\frac{\delta (r)}{r}.
\end{equation}
By expanding the above equation, we arrive at the Laplace-Beltrami operator
in the curved space
\begin{equation}
\nabla _{\alpha }^{2}=\frac{\partial ^{2}}{\partial r^{2}}+\frac{1}{r}\frac{
\partial }{\partial r}+\frac{1}{\alpha ^{2}r^{2}}\frac{\partial ^{2}}{
\partial \varphi ^{2}}.
\end{equation}
In the present system, the total angular momentum is the sum of the angular
momentum and the spin, ${J}=-i\partial/\partial \varphi+s/2$.
Since $J$ commutes
with $H$, we seek solutions of the form
\begin{equation}
\chi_{s}=\sum_{m}\psi_{m}(r)\,e^{im\varphi },  \label{eq:phis}
\end{equation}
with $m=0,\pm 1,\pm 2,\pm 3,\ldots $ being the angular momentum quantum
number and $\psi_{s}(r)$ satisfies the differential equation
\begin{equation}
  h \psi_{m}(r)=k^{2}\psi_{m}(r),
  \label{eq:eigen}
\end{equation}
with
\begin{equation}
  h=h_{0}+\lambda \frac{\delta (r)}{r},
    \label{eq:hfull}
\end{equation}
and
\begin{equation}
h_{0}=-\frac{d^{2}}{dr^{2}}-\frac{1}{r}\frac{d}{dr}+\frac{j^{2}}{r^{2}}.
\label{eq:hzero}
\end{equation}
The parameter $j$ represents the effective angular momentum
\begin{equation}
  j=\frac{m+\phi }{\alpha }-\frac{(1-\alpha )s}{2\alpha},
  \label{eq:eamj}
\end{equation}
and
\begin{equation}
\lambda =\frac{g_{e}\phi s}{2\alpha }.
\end{equation}
By observing equation \eqref{eq:eamj}, one can verify that the presence
of the spin element in the model leads to the appearance of a $\delta$
function potential.
The quantity $\lambda \delta(r)/r$ in Eq. \eqref{eq:hfull} is
interpreted as the interaction between the spin of the particle and the
AB flux tube.
As pointed out by Hagen \cite{PRL.64.503.1990,IJMPA.6.3119.1991} in flat
space ($\alpha=1$), this interaction affects the scattering phase shift.
In this work, by using the self-adjoint extension approach, we shall
confirm these results and show that this delta function also leads
to bound states.
This approach had to be adopted to deal with singular Hamiltonians in
previous works as, for example, in the study of spin 1/2 AB system and
cosmic strings \cite{EPJC.74.2708.2014,PRD.40.1346.1989}, in the
Aharonov-Bohm-Coulomb problem
\cite{EPJC.73.2548.2013,TMP.161.1503.2009,PRD.50.7715.1994,
JMP.36.5453.1995},
and the study of the equivalence between the self--adjoint extension
method and renormalization \cite{Book.1995.Jackiw}.

\subsection{Application of the BG method}
\label{subsec:app_BG}

In this section, we employ the KS method to find the S-matrix and from
its poles we obtain an expression for the bound states.
To apply the BG method, we need first transform the operator $h_0$ in
\eqref{eq:hzero} to compare with the form in Eq. \eqref{eq:hBG}.
This is accomplished by employing a similarity transformation by means
of the unitary operator
$U:L^{2}(\mathbb{R}^{+},rdr)\rightarrow L^{2}(\mathbb{R}^{+},dr)$, given
by $(U\xi)(r)=r^{1/2}\xi(r)$.
Thus, the operator $h_{0}$ becomes
\begin{equation}
  \tilde{h}_{0}=UH_{0}U^{-1}=
  -\frac{d^{2}}{dr^{2}}+
    \left(j^2-\frac{1}{4}\right)
    \frac{1}{r^{2}},
\end{equation}
and by comparing with \eqref{eq:hBG} we must have $\gamma=\beta=W=0$ and
\begin{equation}
\ell(\ell-1)=j^2-\frac{1}{4}.
\end{equation}
It is well-known that the radial operator $h_{0}$ is not
essentially self-adjoint
for $\ell(\ell-1)<3/4$, otherwise it is essentially self-adjoint
\cite{Book.1975.Reed.II}.
Therefore, using the above equation in this inequality, we have
\begin{equation}
  |j|<1.
  \label{eq:jineq}
\end{equation}

\begin{figure*}
  \includegraphics[width=0.47\textwidth]{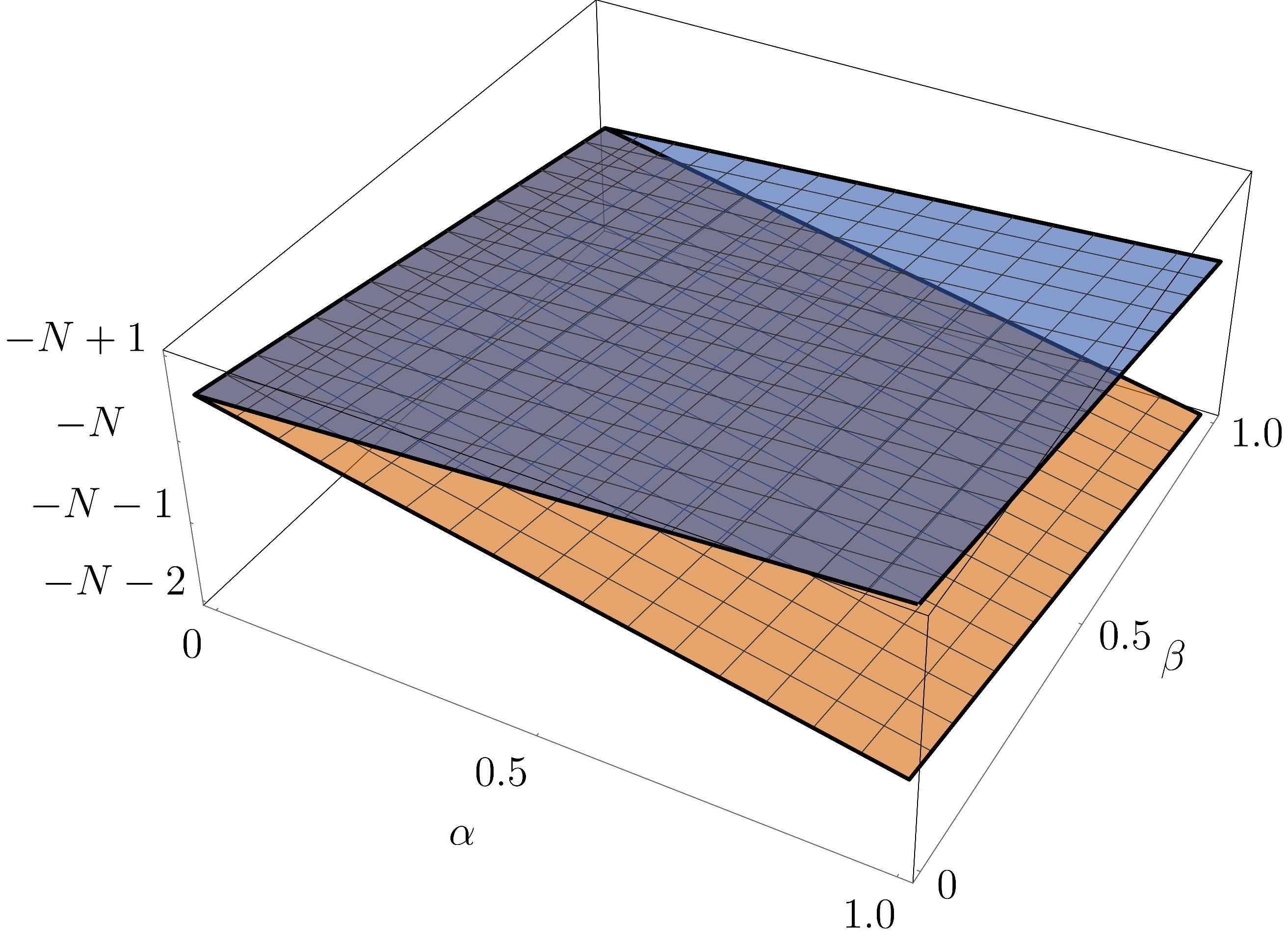}
  \hspace{0.5cm}
  \includegraphics[width=0.47\textwidth]{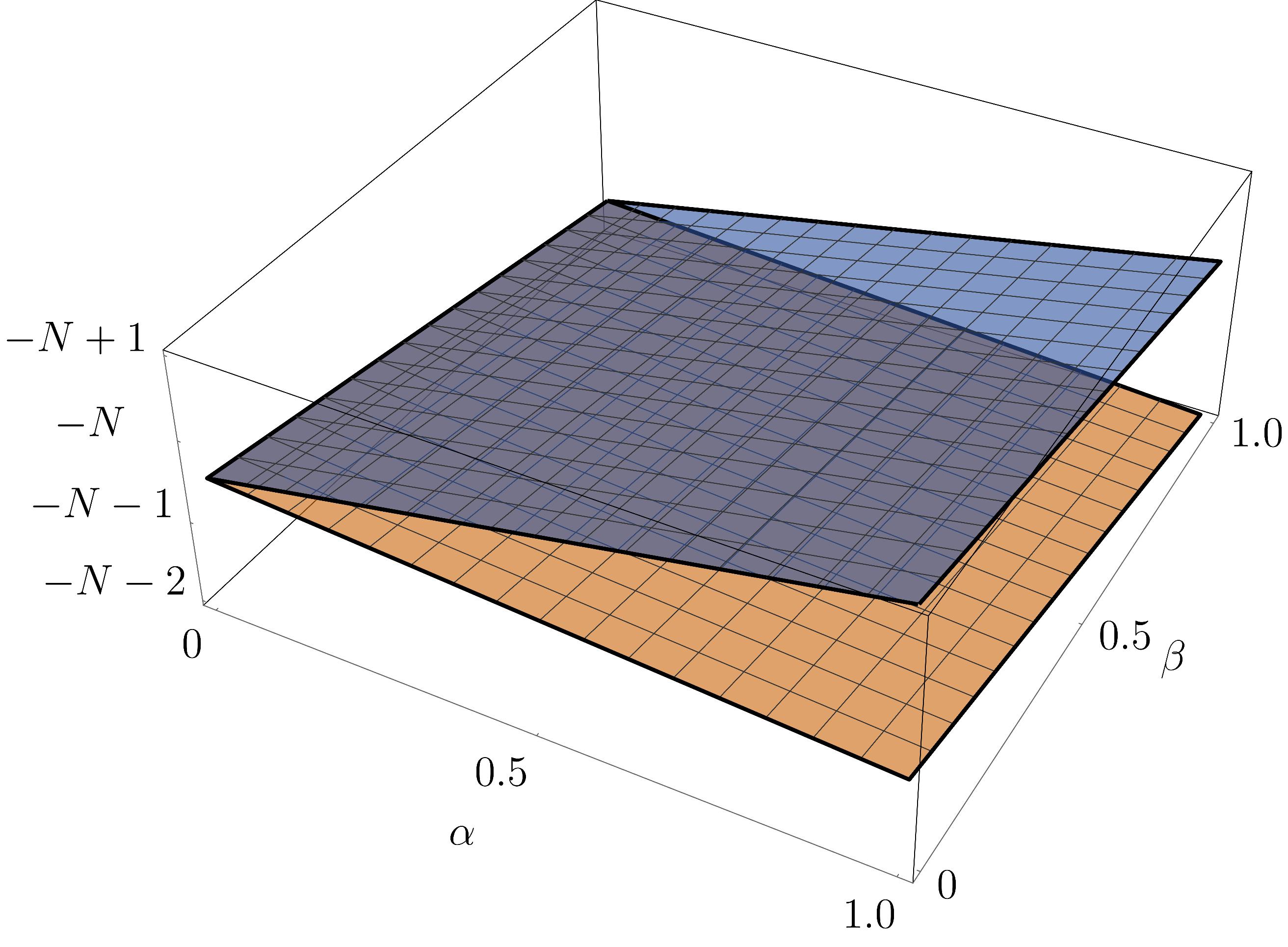}\\
  \includegraphics[width=0.47\textwidth]{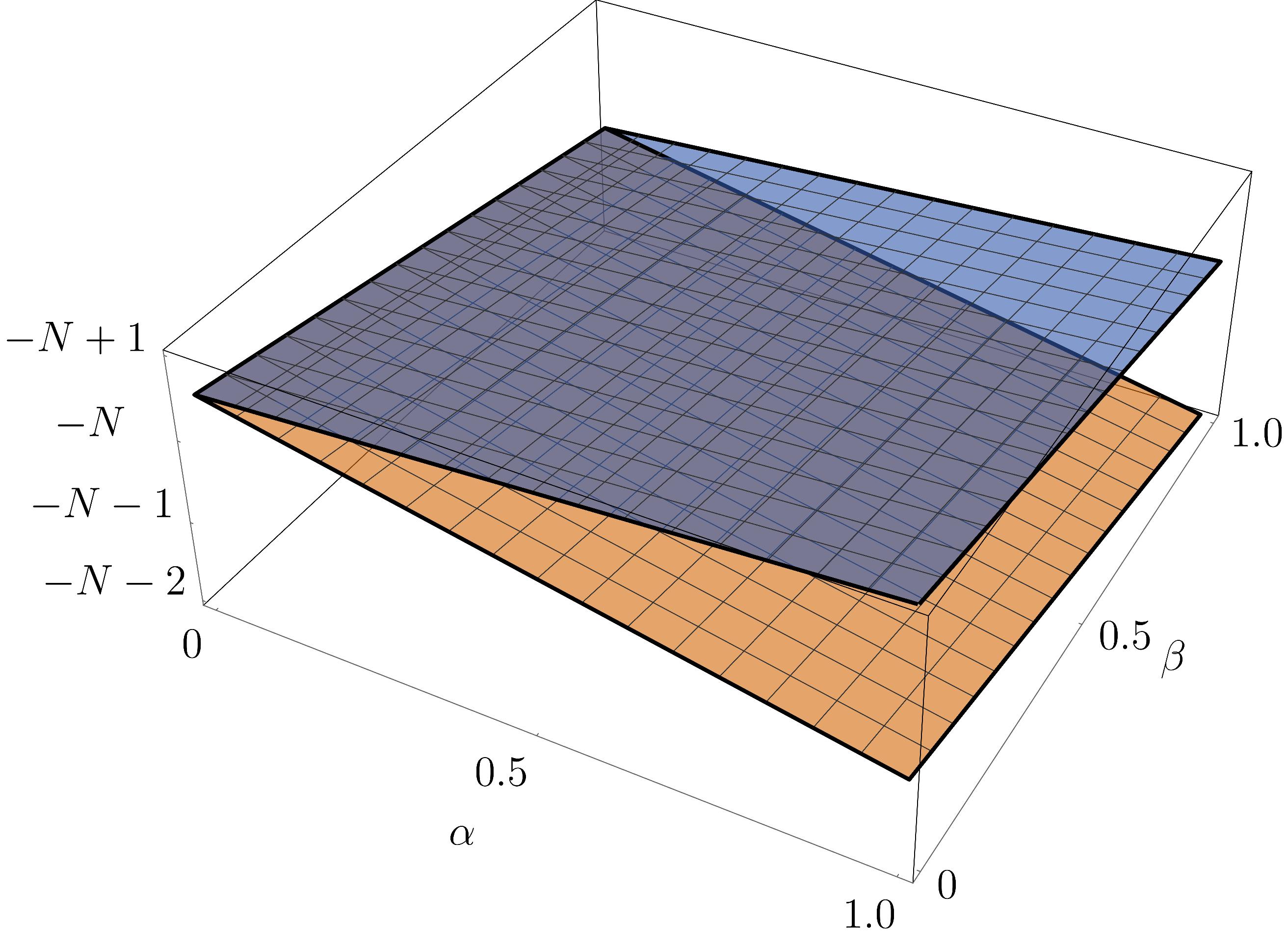}
  \hspace{0.5cm}
  \includegraphics[width=0.47\textwidth]{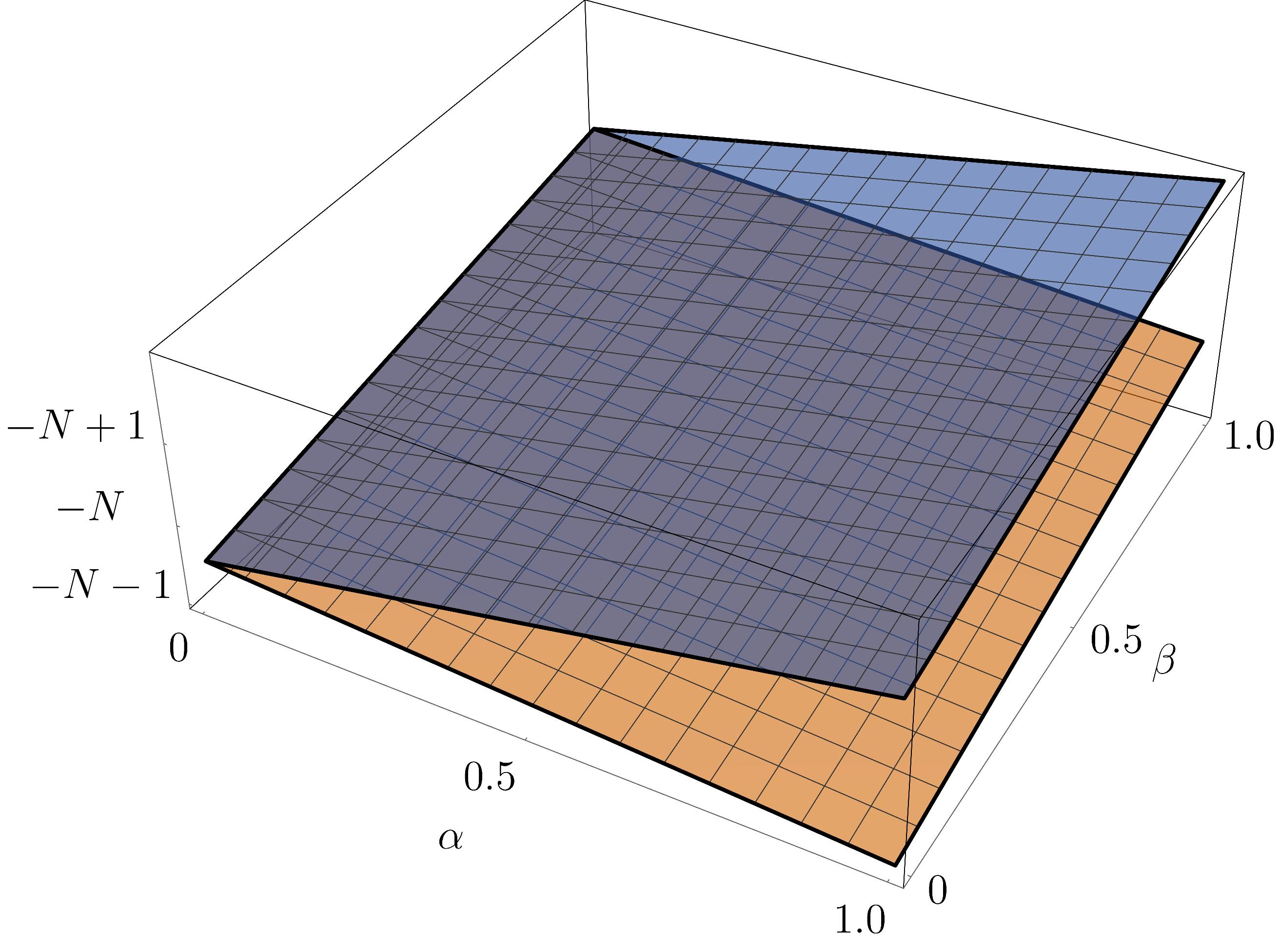}
  \caption{
    In this figure we show the graphs of the planes
    $\pi_{\pm}^{AB}(\alpha,\beta)$ for the AB (top panel) and the planes
    $\pi_{\pm}^{AC}(\alpha,\beta)$ fot the AC (bottom panel) effects.
    The figures on the left are for $s=-1$ and on the right is for
    $s=+1$.
    The planes delimit the region where $h_0$ is not self-adjoint.}
  \label{fig:fig1}
\end{figure*}

\begin{figure*}
  \centering
  \includegraphics[width=0.47\textwidth]{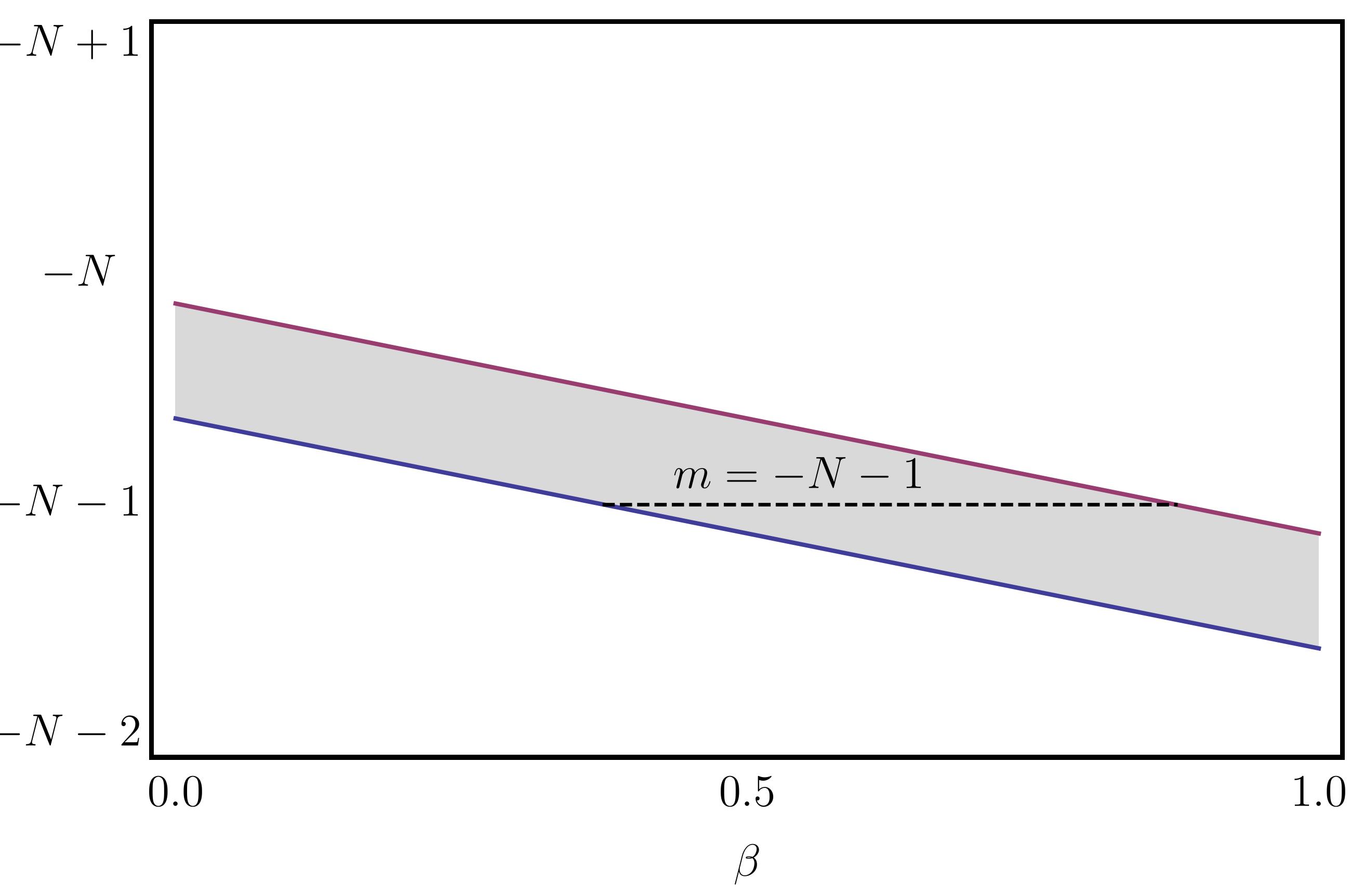}
  \hspace{0.5cm}
  \includegraphics[width=0.47\textwidth]{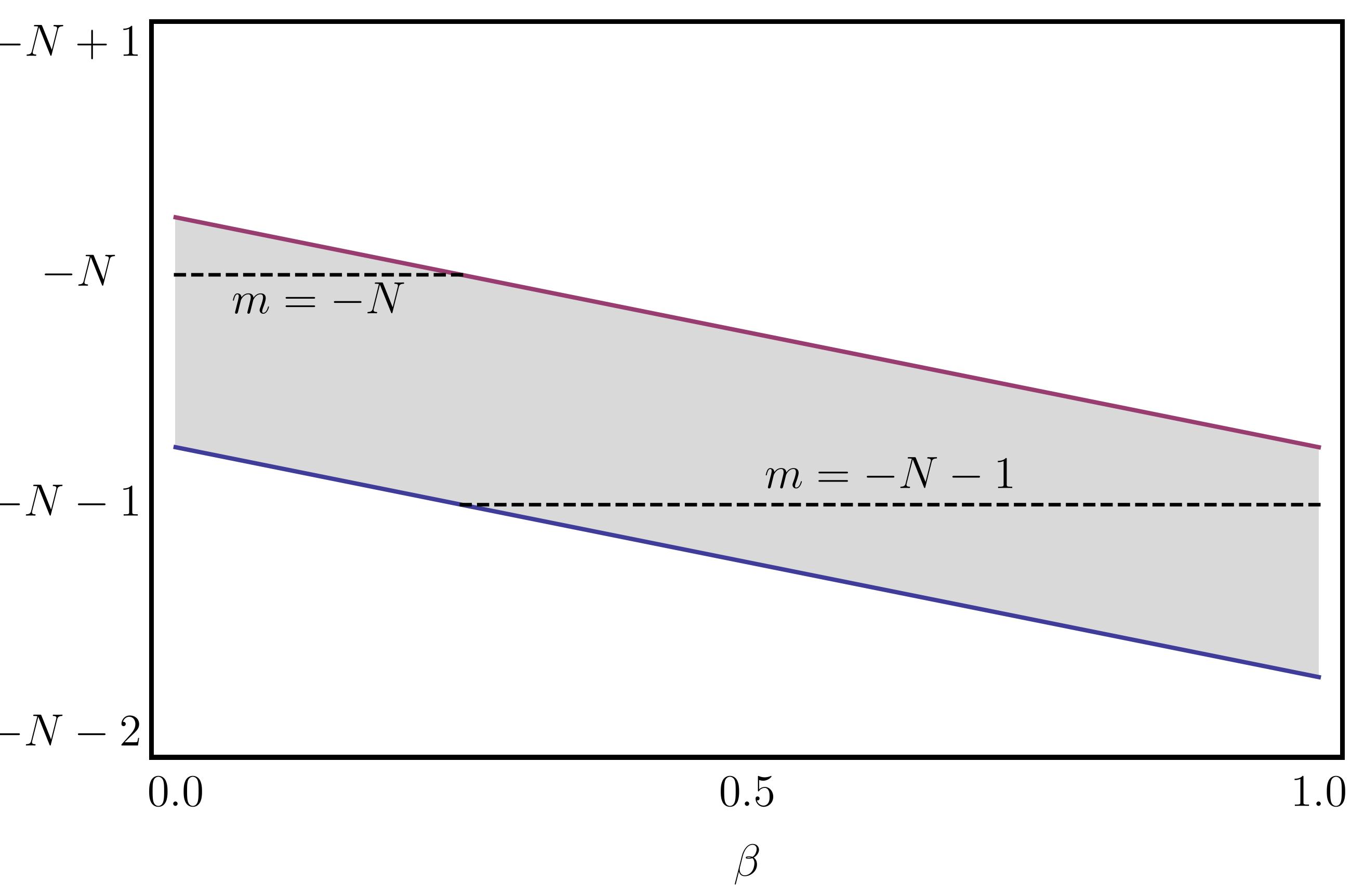}\\
  \includegraphics[width=0.47\textwidth]{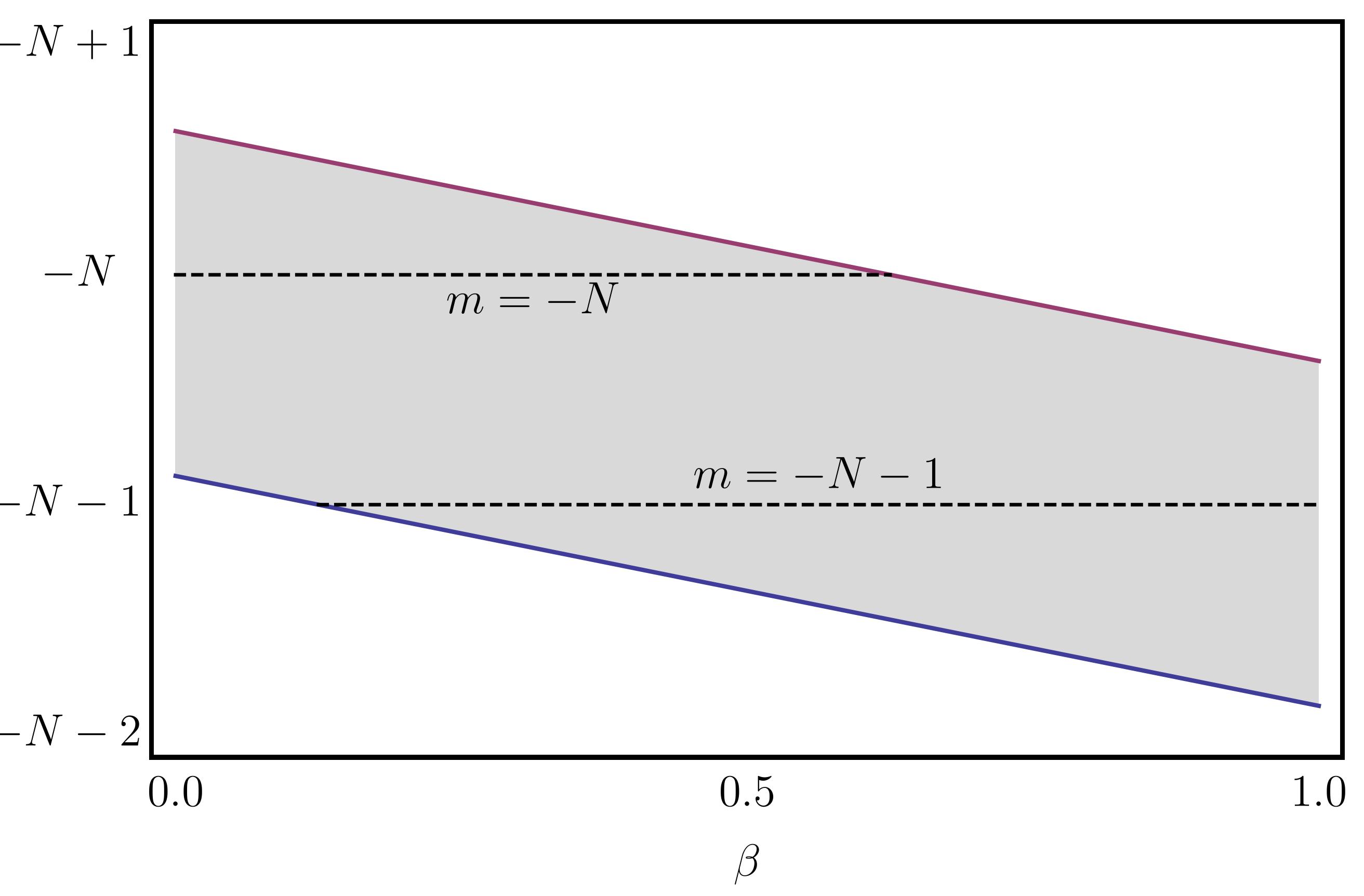}
  \hspace{0.5cm}
  \includegraphics[width=0.47\textwidth]{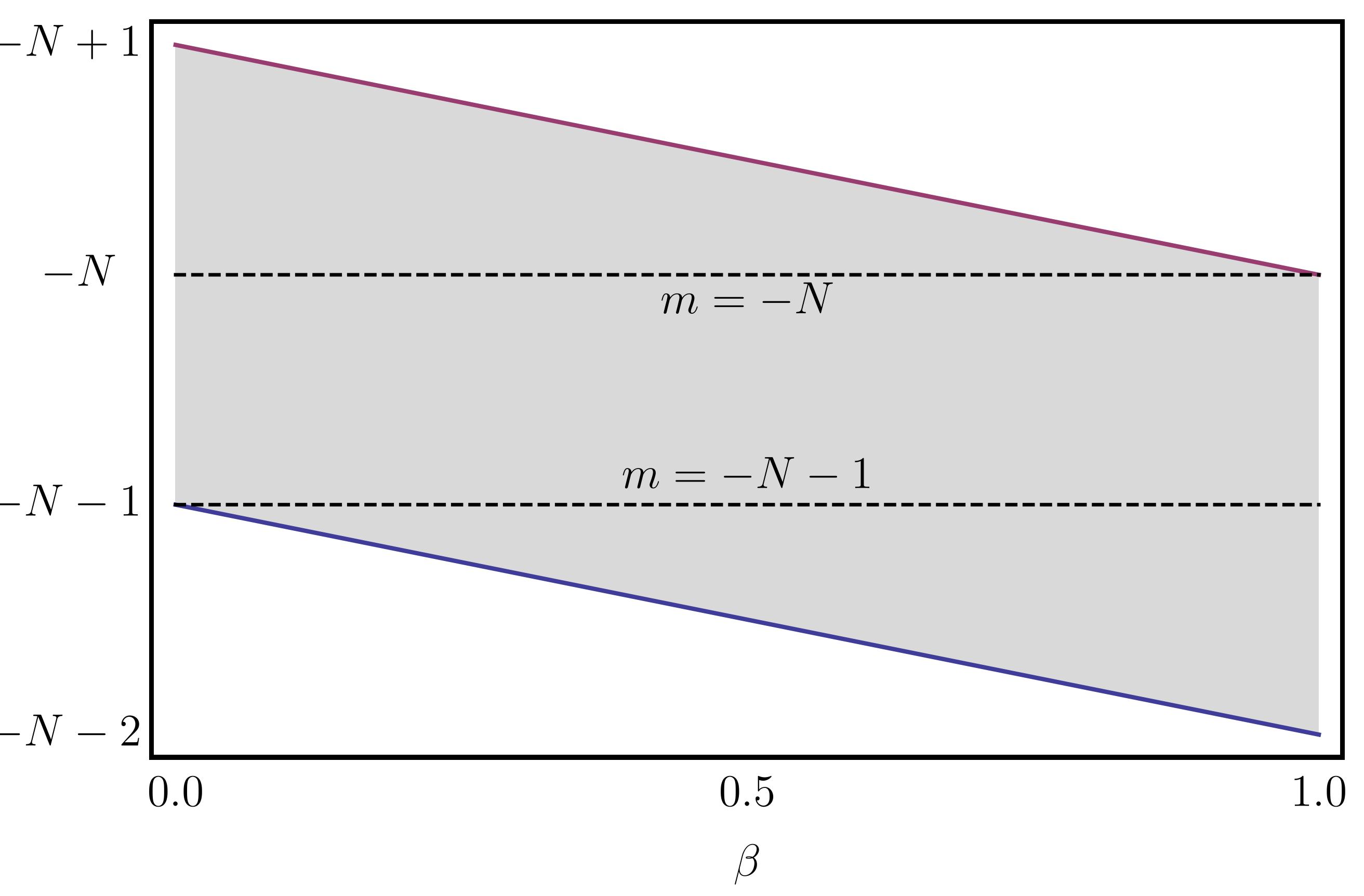}
  \caption{Cross sections of Fig. \ref{fig:fig1} (top left panel) with
    $s=-1$ for:
    $\alpha$ = 0.25 (top left panel), $\alpha$ = 0.50 (top right panel),
    $\alpha$ = 0.75 (bottom left panel), and $\alpha$ = 1 (bottom right
    panel).
    The area of the stripe detached in the figure represents the region
    in which the operator $h_0$ is not self-adjoint.
    The dashed lines refer to the values of angular momentum
    quantum number.}
  \label{fig:fig2}
\end{figure*}

\begin{figure*}
  \includegraphics[width=0.47\textwidth]{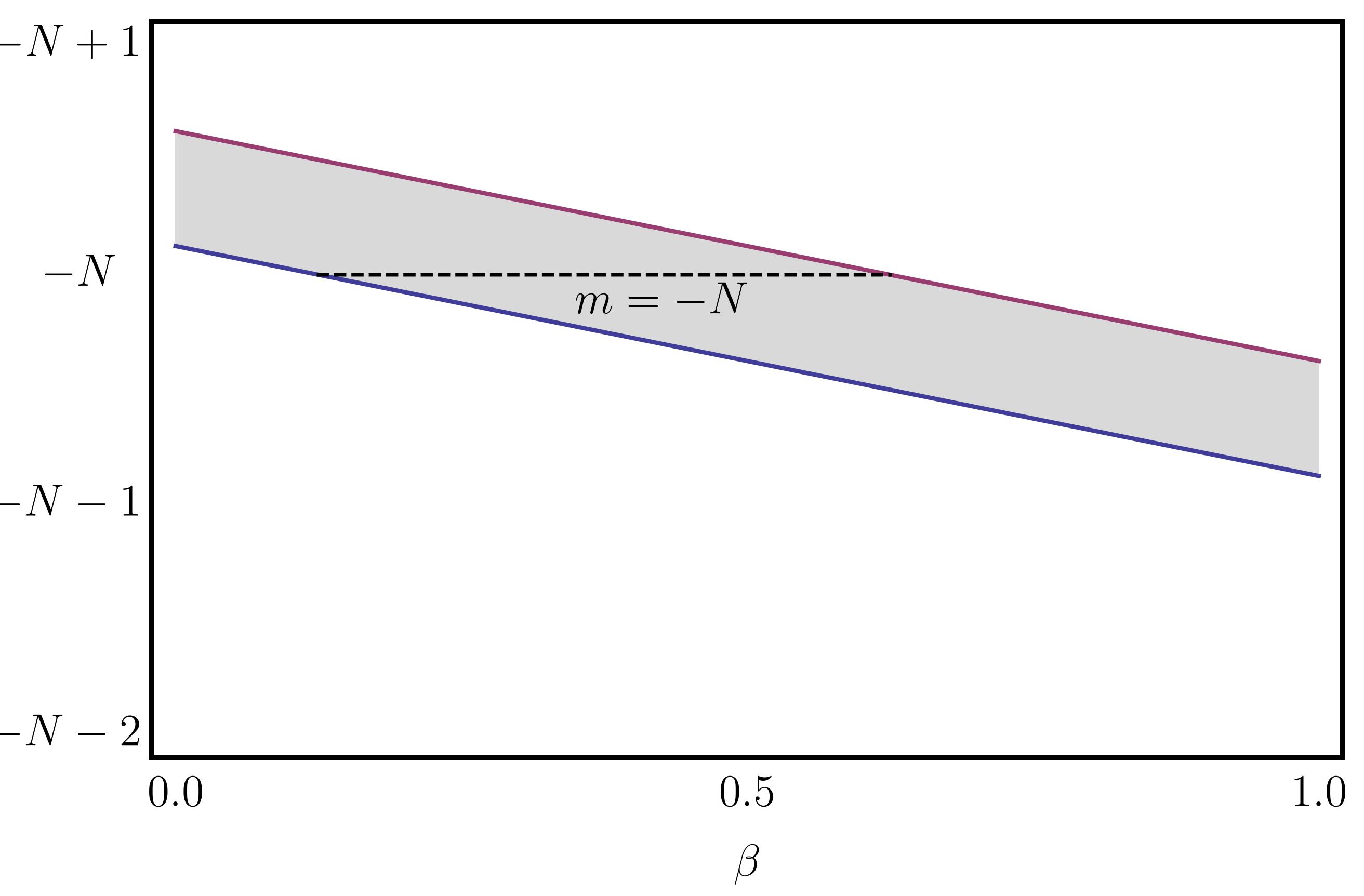}
  \hspace{0.5cm}
  \includegraphics[width=0.47\textwidth]{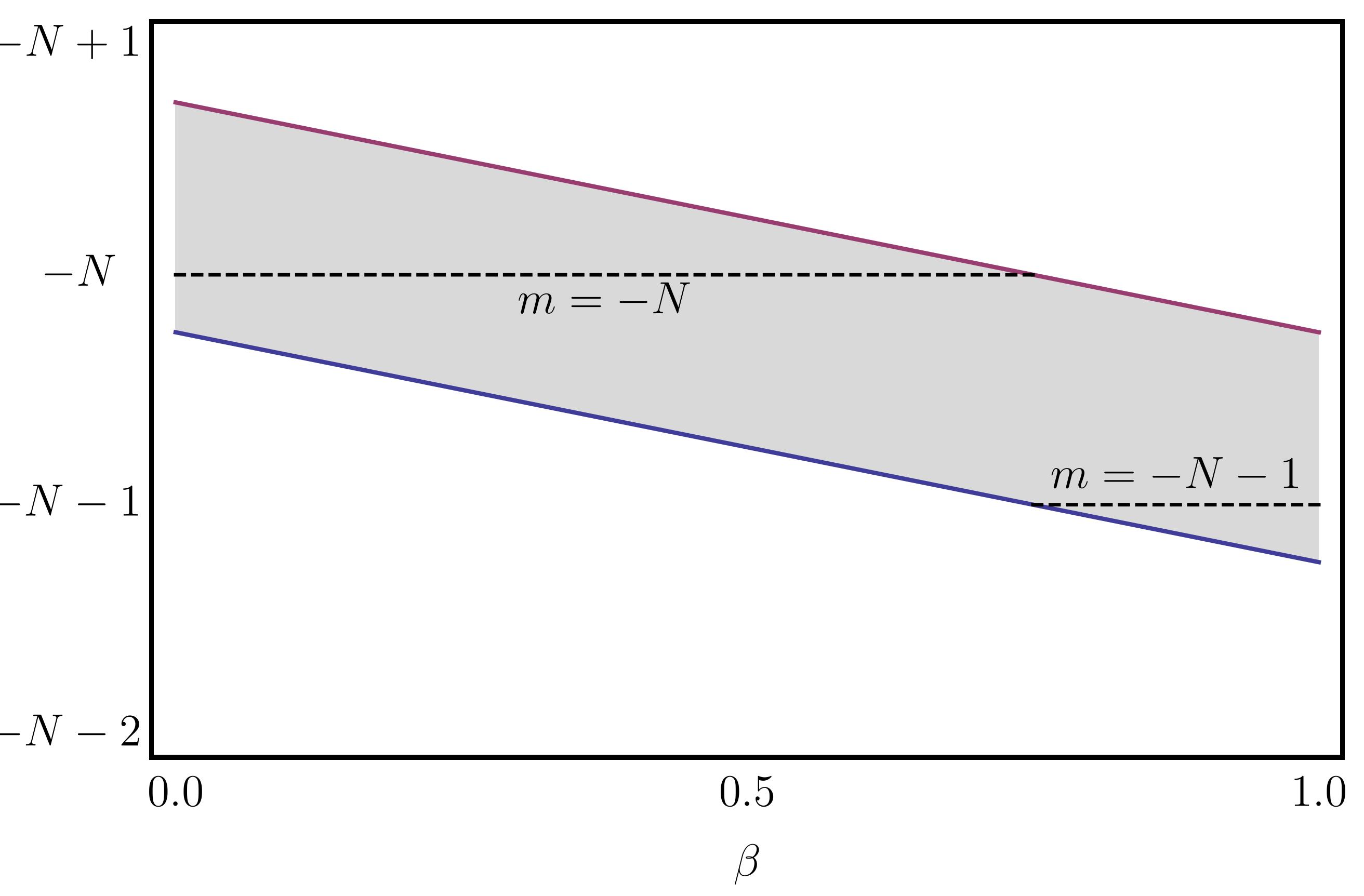}\\
  \includegraphics[width=0.47\textwidth]{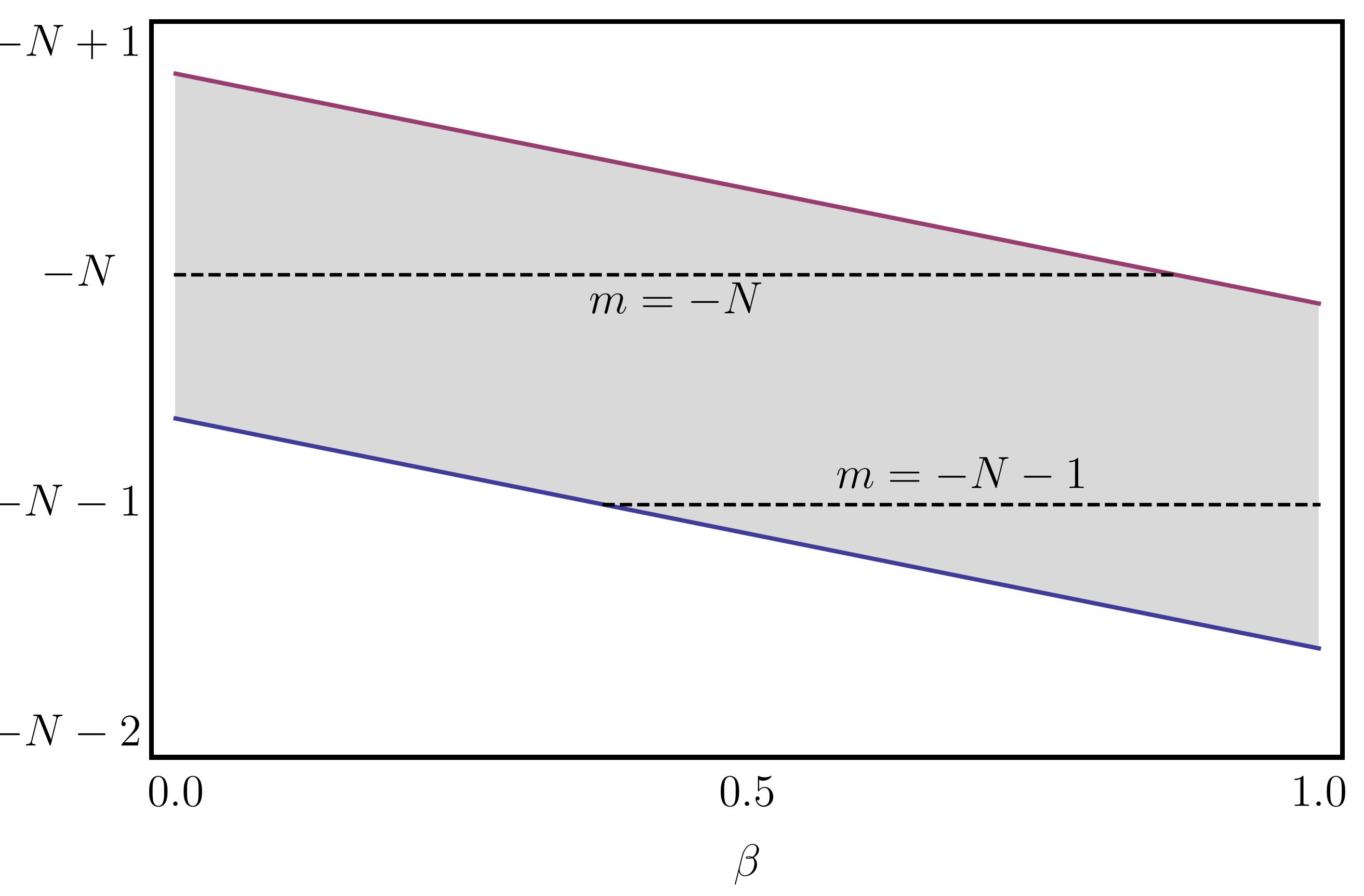}
  \hspace{0.5cm}
  \includegraphics[width=0.47\textwidth]{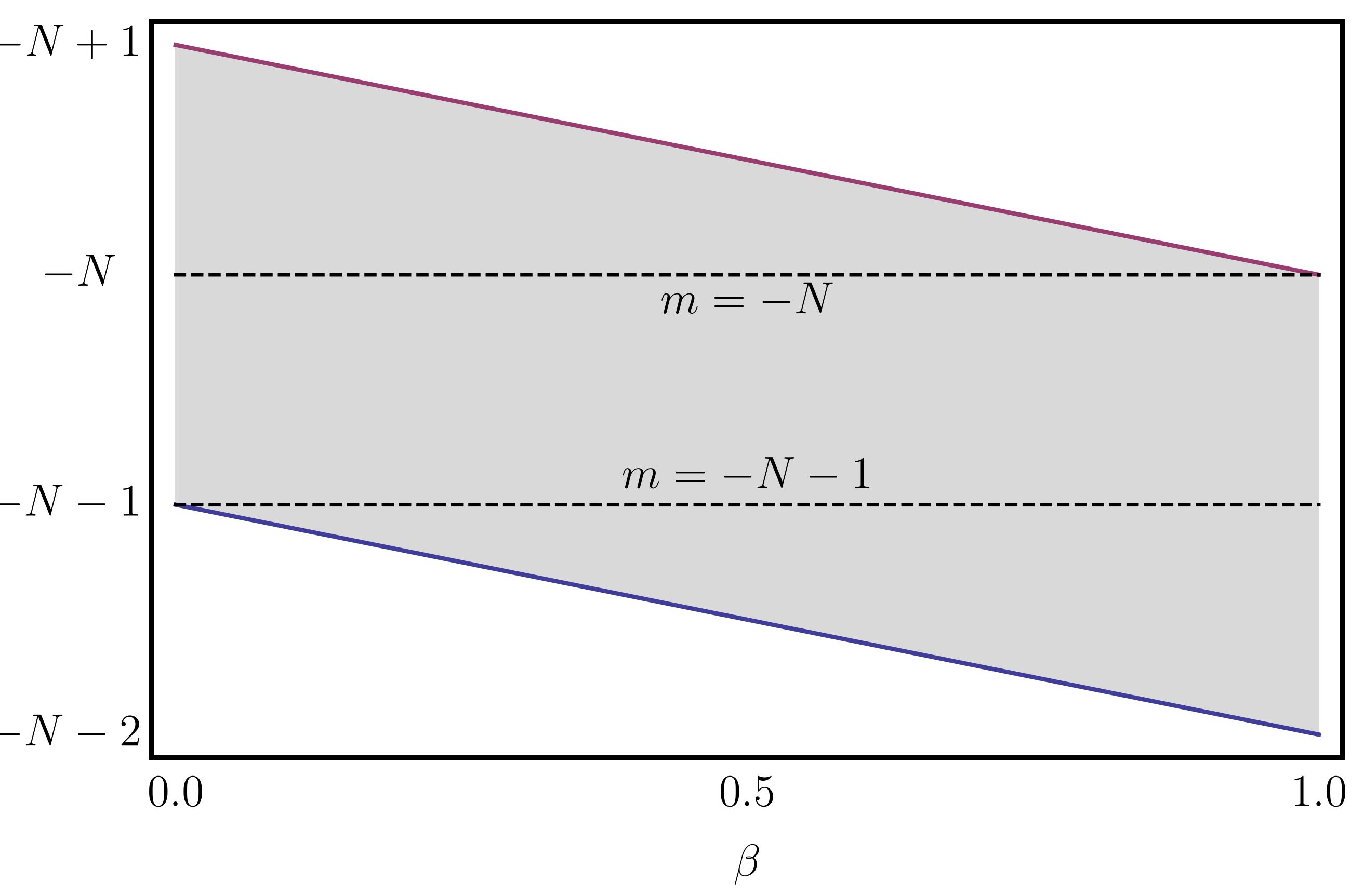}
  \caption{
    Cross sections of Fig. \ref{fig:fig1} (top left panel) with
    $s=+1$ for:
    $\alpha$ = 0.25 (top left panel), $\alpha$ = 0.50 (top right panel),
    $\alpha$ = 0.75 (bottom left panel), and $\alpha$ = 1 (bottom right
    panel).
    The area of the stripe detached in the figure represents the region
    in which the operator $h_0$ is not self-adjoint.
    The dashed lines refer to the values of angular momentum quantum
    number.
  }
  \label{fig:fig3}
\end{figure*}

Before we going to the application of Theorem \ref{th:BG}, it is
interesting to get a deeper understanding of the significance of
the above equation for it informs us for which values of the angular
momentum quantum number $m$ the operator  $h_0$ is not self-adjoint.
From Eq. \eqref{eq:eamj}, we see that these values are dependent on the
magnetic quantum flux $\phi$, the value of $\alpha$ and the spin
parameter $s$.
By employing the decomposition of the magnetic quantum
flux as
\begin{equation}
  \phi = N + \beta,
\end{equation}
with $N$ being the largest integer contained in $\phi$ and
\begin{equation}
  0 \leq \beta < 1,
\end{equation}
the inequality in Eq. \eqref{eq:jineq}, becomes
\begin{equation}
  \pi_{-}^{\rm AB} (\alpha,\beta)< m < \pi_{+}^{\rm AB} (\alpha,\beta),
\end{equation}
with
\begin{equation}
  \pi_{\pm}^{\rm AB}(\alpha,\beta)=\pm\alpha-(N+\beta)
  +\frac{(1-\alpha)s}{2}.
  \label{eq:ABplanes}
\end{equation}
The planes $\pi_{\pm}^{\rm AB}(\alpha,\beta)$ delimit the region in
which $h_0$ is not self-adjoint.
Given the exact equivalence of the spin 1/2 AB and AC effects
\cite{PRL.64.2347.1990}, Eq.  \eqref{eq:ABplanes} should be compared
with the corresponding planes obtained for the AC effect in the conical
space.
In Ref. \cite{AHEP.2017.2017.7} it was shown that the planes for the AC
effect are given by
\footnote{there is a missprint in the signal of the term $sN$ in
$\pi_{\pm}(\alpha,\beta)$ in Ref. \cite{AHEP.2017.2017.7}}
\begin{equation}
  \pi_{\pm}^{\rm AC}(\alpha,\beta)=\pm\alpha-s(N+\beta)
  +\frac{(1-\alpha)s}{2}.
\end{equation}
Although the equations for the planes are very similar, there is an
additional dependence on the spin parameter $s$ in the AC effect.
In Fig. \ref{fig:fig1} we show the planes for AB (top panel) and AC
(bottom panel) effects as a function of $\beta$ and it is possible to
see in the AB effect the $s$ parameter changes the values of $m$ in
which $h_0$ is not self-adjoint and the planes are decreasing functions
of $\beta$ whatever the value of $s$ while in the AC effect, besides of
changing the values of $m$, it also controls the inclination of the
planes (compare the figures at the bottom panel of Fig. \ref{fig:fig1}.
We can have even more information about the affected $m$ values
(in the sense of which values of it $h_0$ is not self-adjoint)
by looking at some specific values of $\alpha$.
Thus, in Fig. \ref{fig:fig2} and \ref{fig:fig3} we show cross
sections of Fig. \ref{fig:fig1} for $s=-1$ and $s=+1$, respectively.
In Fig. \ref{fig:fig2} (\ref{fig:fig3}) we can see that for
$s=-1$ ($s=+1$) and $\alpha=0.25$ only for $m=-N-1$ ($m=-N$) the
operator $h_0$ is not self-adjoint.
On the other hand, for $\alpha=0.50$ for both values of $m=-N$ and
$m=-N-1$ the operator $h_0$ is not self-ajoint.
In fact, the minimum value of $\alpha$ in which $h_0$ is not
self-adjoint for both values of $m$ is $\alpha_{\rm min}= 1/3$.
Moreover, for $\alpha=1$ (flat space), the operator $h_0$ is not
self-adjoint for both values of angular momentum for all range of
$\beta$, which is a very well-known result
\cite{Book.2004.Albeverio,LMP.43.43.1998,JMP.39.47.1998,AoP.146.1.1983}.

Now that we have discussed in detail the significance of inequality
$|j|<1$, we can return to our main discussion.
Thus, in the subspace where $|j|<1$, we must apply Theorem \ref{th:BG}, in
such a way that all the  self--adjoint extensions  $h_{0,\nu}$ of
$h_{0}$ are characterized by the boundary condition at the origin
\begin{equation}
  \nu \psi_{0,j} = \psi_{1,j},
  \label{eq:bc}
\end{equation}
with $-\infty < \nu \leq \infty$, $-1 < j < 1$ and the boundary values
are
\begin{align*}
  \psi_{0,j} = {}
  & \lim_{r \to 0^{+}}r^{|j|}\psi_{m}(r),  \nonumber \\
  \psi_{1,j} = {}
  & \lim_{r \to 0^{+}}\frac{1}{r^{|j|}}
    \left[
    \psi_{m}(r)-\psi_{0,j}\frac{1}{r^{|j|}}\right].
\end{align*}
Physically, it turns out that we can interpret $1/\nu$ as the scattering
length of $h_{0,\nu}$ \cite{Book.2004.Albeverio}.
For $\nu=\infty$ (the Friedrichs extension of $h_{0}$), we obtain the
free Hamiltonian (the case describing spinless particles) with regular
wave functions at the origin ($\psi_{m}(0)=0$).
This scenario is similar to imposing the Dirichlet boundary condition
on the wave function and recovers the original result of Aharanov and
Bohm \cite{PR.115.485.1959}.
On the other hand, if $|\nu|<\infty$, $h_{0,\nu}$ characterizes a point
interaction at $r=0$ and the boundary condition permits a $r^{-|j|}$
singularity in the wave functions at this point
\cite{PRA.77.036101.2008}.

Now that we have a suitable boundary condition, we can return to Eq.
\eqref{eq:eigen} and look for its solutions.
Equation \eqref{eq:eigen} is nothing more than the Bessel differential
equation for $r \neq 0$.
Thus, the general solution for $r\neq 0$ is given by
\begin{equation}
 \psi_{m}(r)=a_{m}J_{|j|}(kr)+b_{m}J_{-|j|}(kr),  \label{eq:Bessel}
\end{equation}
where $J_{\nu}(z)$ is the Bessel function of fractional order and
$a_{m}$ and $b_{m}$ are the coefficients corresponding to the
contributions of the regular and irregular solutions at $r=0$,
respectively.
By means of the boundary condition in Eq. \eqref{eq:bc}, we obtain a
relation between $a_{m}$ and $b_m$,
\begin{equation}
  b_{m}=-\mu_{\nu} a_{m},
  \label{eq:relation}
\end{equation}
which is valid in the subspace $|j|<1$.
The term $\mu_{\nu}$ is given by
\begin{equation}
  \mu_{\nu} =
   \frac
  {k^{2|j|}\Gamma(1-|j|)\sin(|j| \pi)}
  {4^{|j|}\Gamma(1+|j|) \nu+ k^{2|j|}\Gamma(1-|j|)\cos(|j|\pi)},
  \label{eq:mu}
\end{equation}
where $\Gamma(z)$ is the gamma function.
In Eq. \eqref{eq:mu} one can verify that $\mu_{\nu}$ controls, through $\nu$, the contribution of the irregular solution $J_{-|j|}$ for the
wave function.
Thus, the solution in this subspace reads
\begin{equation}
  \psi_{m}(r)=a_{m}\left[J_{|j|}(kr)-\mu_{\nu} J_{-|j|}(kr)\right].
\end{equation}
We can observe that for $\nu=\infty$, we obtain $\mu_{\infty}=0$ and, in
this case, there is no contribution of the irregular solution at the
origin for the wave function.
Consequently, in this case, the total wave function becomes
\begin{equation}
  \psi = \sum_{m=-\infty}^{\infty} a_m J_{|j|}(k r) e^{i m \varphi}.
  \label{eq:regpsi}
\end{equation}
The coefficient $a_{m}$ in Eq. \eqref{eq:regpsi} must be chosen in such
a way that $\psi$ represents a plane wave that is incident from the
right.
In this case, we find the following result:
\begin{equation}
  a_{m} = e^{-i|j| \pi/2}.
\end{equation}
The scattering phase shift can be obtained from the asymptotic behavior
of Eq. \eqref{eq:regpsi}.
This leads to
\begin{equation}
  \delta_{m}= \frac{\pi}{2}(|m|-|j|).
  \label{eq:regphaseshift}
\end{equation}
This is the scattering phase shift of the AB effect in the cosmic string
spacetime \cite{PRD.85.041701.2012,AoP.339.510.2013}.
It is important to mention that, for $\alpha=1$, it reduces to the phase
shift for the usual AB effect in flat space
$\delta_{m}^{AB}=\pi(|m|-|m+\phi|)/2$ \cite{PR.115.485.1959}.

On the other hand, for $|\nu| < \infty$, the contribution of the
irregular solution changes the scattering phase shift to
\begin{equation}
  \delta_{m}^{\nu}=
  \delta_{m}+\arctan (\mu_{\nu}).
 \label{eq:phaseshift}
\end{equation}
Thus, from standard results for the S-matrix, one obtains
\begin{align}
  S_{m}^{\nu}
  =
  e^{2 i \delta_{m}^{\nu}}
  =
  e^{2 i \delta_{m}}
  \left(
  \frac{1+i\mu_{\nu}}{1-i\mu_{\nu}}
  \right),
  \label{eq:Sm}
\end{align}
which is the expression for the S-matrix given in terms of phase shift.
It can be seen in \eqref{eq:Sm} that there is an additional scattering
for any value of the self--adjoint extension parameter $\nu$.
By choosing $\nu=\infty$, we find the S-matrix for the AB effect in the
cosmic string spacetime, as it should be.

Having obtained the S-matrix, the bound state energies can be identified
as the poles of it in the upper half of the complex $k$ plane.
To find them, we need to examine the zeros of the denominator
in Eq. \eqref{eq:Sm}, $1-i\mu_{\nu}$, with the replacement
$k\rightarrow i\kappa_{b}$ with $\kappa_{b}=\sqrt{2ME_{b}}$.
Therefore, for $\nu<0$, the bound state energy is given by
\begin{equation}
  E_{b}=-\frac{2}{M}
  \left[-\nu \frac{\Gamma(1+|j|)}{\Gamma(1-|j|)}\right]^{1/|j|}.
\label{eq:energy_BG}
\end{equation}
Thus, for a fixed negative value of the self-adjoint extension parameter
$\nu$, there is a single bound state and the value $2|\nu|^{1/|j|}/M$
fixes the energy scale.
The result in Eq. \eqref{eq:energy_BG} coincides with the bound state
energy found in Refs. \cite{PRD.85.041701.2012,AoP.339.510.2013} for the
AB effect in curved space and is similar that one found in contact
interactions of anyons \cite{PLB.268.222.1991}.
It is also possible to express the S-matrix in terms of the bound
state energy.
The result is seen to be
\begin{equation}
  S_{m}^{\nu} = e^{2i\delta_{m}}
  \left[
    \frac
    {e^{ 2 i \pi |j|}-(\kappa_{b}/k)^{2 |j|}}
    {1-(\kappa_{b}/k)^{2 |j|}}
  \right].
\end{equation}

It is important to comment that the above results for the scattering
matrix and the bound state energy (for $\nu<0$) are valid only when
$|j|<1$.
Moreover, all the results are dependent on a free parameter,
the self-adjoint extension parameter $\nu$.
In what follows we shall show that by employing the KS method, we can
find an expression relating the self-adjoint extension parameter with
physical parameters of the system.

\subsection{Application of the KS method}
\label{subsec:app_KS}

In this section, we employ the KS approach to find the bound states for
the Hamiltonian in Eq. \eqref{eq:hfull}.
Following the discussion in Sec. \ref{subsec:kay_studer}, we temporarily
forget the $\delta$-function potential in $h$ and substitute the problem
in Eq. \eqref{eq:eigen} by the
eigenvalue equation for $h_{0}$,
\begin{equation}
  h_{0}\psi_{\rho}=k^{2}\psi_{\rho},
\label{eq:ideal}
\end{equation}
plus self-adjoint extensions.
Here, $\psi_{\rho}$ is labelled by the parameter $\rho$ of the
self-adjoint extension, which is related to the behaviour of the wave
function at the origin.
To turn $h_{0}$ into a self-adjoint operator its domain of definition
has to be extended by the deficiency subspace, which is spanned by the
solutions of the eigenvalue equation (cf. Eq. \eqref{eq:defspaces}
\begin{equation}
h_{0}^{\dagger}\psi_{\pm}=\pm i k_{0}^{2} \psi_{\pm},
\label{eq:eigendefs}
\end{equation}
where $k_{0}^{2}\in \mathbb{R}$ is introduced for dimensional
reasons.
Since $h_0$ is Hermitian, $h_{0}^{\dagger}=h_{0}$, the only
square integrable functions which are solutions of
Eq. \eqref{eq:eigendefs} are the modified Bessel functions of
second kind,
\begin{equation}
\psi_{\pm}=K_{|j|}( \sqrt{\mp i}k_{0}r),
\end{equation}
with $\Im \sqrt{\pm i}>0$.
These functions are square integrable only in the range
$j\in(-1,1)$, for which $h_{0}$ is not self-adjoint.
The dimension of such deficiency subspace is thus $(n_{+},n_{-})=(1,1)$,
in agreement with the results of the previous sections.
In this manner, $\mathcal{D}(h_{\rho,0})$ in $L^{2}(\mathbb{R}^{+},rdr)$
is given by the set of functions \cite{Book.1975.Reed.II}
\begin{equation}
  \psi_{\rho}(r)=\psi_{m}(r)+C
  \left[ K_{|j|}(\sqrt{-i}k_{0}r)+
    e^{i\rho}K_{|j|}(\sqrt{i}k_{0}r)
  \right],
  \label{eq:domain}
\end{equation}
where $\psi_{m}(r)$, with $\psi_{m}(0)=\dot{\psi}_{m}(0)=0$, is the
regular wave function and the
mathematical parameter $\rho\in [0,2\pi)$ represents a choice for the
boundary condition.
For different values of $\rho$, we have different domains for $h_0$.
and the adequate boundary condition will be determined by the physical
system.
\cite{AoP.325.2529.2010,AoP.323.3150.2008,PRD.40.1346.1989,
  JMP.53.122106.2012}.
Thus, in this direction, we use a physically motivated regularization for
the magnetic field.
So, we replace the original potential vector of the AB flux tube by the
following one
\cite{PRL.64.503.1990,PRL.64.2347.1990,IJMPA.6.3119.1991,
PRD.48.5935.1993}
\begin{equation}
  e\mathbf{A}=
    \begin{cases}
      \displaystyle
      -\frac{\phi}{ \alpha r}
      \mathbf{\hat{\varphi}}, &r>r_{0},\\
      0, & r<r_{0}.
  \end{cases}
\end{equation}
where $r_0$ is a length that defines the defect core radius
\cite{CMP.139.103.1991,AoP.323.3150.2008}, which is a very small radius
smaller than the Compton wave length $\lambda_C$ of the electron \cite{PLB.333.238.1994}.
So one makes the replacement
\begin{equation}
\frac{\delta(r)}{r} \to \frac{\delta(r-r_{0})}{r_{0}}.
\label{eq:ushortr0}
\end{equation}
This regularized form for the delta function allows us to determine an
expression for $\rho$.
To do so, we consider the zero-energy solutions $\psi_{0}$ and
$\psi_{\rho,0}$ for $h$ with the regularization in \eqref{eq:ushortr0}
and $h_{0}$, respectively,
\begin{equation}
  \left[
    -\frac{d^{2}}{dr^{2}}
    -\frac{1}{r}\frac{d}{dr}
    +\frac{j^2}{r^{2}}
    +\lambda \frac{\delta(r-r_{0})}{r_{0}}
  \right]
  \psi_{0}=0,
\label{eq:statictrue}
\end{equation}
\begin{equation}
  \left[
    -\frac{d^{2}}{dr^{2}}
    -\frac{1}{r}\frac{d}{dr}
    +\frac{j^{2}}{r^{2}}
  \right]
  \psi_{\rho,0}=0.
\label{eq:rhostatic}
\end{equation}
The value of $\rho$ is determined by the boundary condition
\begin{equation}
\lim_{r_{0}\to 0^{+}}r_{0}\frac{\dot{\psi}_{0}}{\psi_{0}}\Big|_{r=r_{0}}=
\lim_{r_{0}\to 0^{+}}r_{0}\frac{\dot{\psi}_{\rho,0}}{\psi_{\rho,0}}\Big|_{r=r_{0}}.
\label{eq:logder}
\end{equation}

The left-hand side of Eq. \eqref{eq:logder} can be obtained by the direct
integration of \eqref{eq:statictrue} from $0$ to $r_{0}$.
The result seems to be
\begin{equation}
\lim_{r_{0}\to 0^{+}}r_{0}\frac{\dot{\psi}_{0}}{\psi_{0}}\Big|_{r=r_{0}}
=\lambda.
 \label{eq:nrs}
\end{equation}

The right-hand side of Eq. \eqref{eq:logder} is calculated as follows.
First, we seek the solutions of the bound states for the Hamiltonian $h_0$.
These solutions will allow us to obtain the solutions of the bound states for
$h$.
As before, for the bound state, we consider $k$ as a pure imaginary
quantity, $k \to i\kappa_{b}$.
So, we have
\begin{equation}
  \left[
    \frac{d^{2}}{dr^{2}}+\frac{1}{r}\frac{d}{dr}-
    \left(\frac{j^{2}}{r^{2}}+
      \kappa_{b}^{2}\right)
  \right] \psi_{\rho}(r)=0,
  \label{eq:eigenvalue}
\end{equation}
The solution for the above equation is the modified Bessel functions
\begin{equation}
  \psi_{\rho}(r)=K_{|j|}
  \left(\kappa_{b} r\right).
\label{eq:sver}
\end{equation}
Second, we observe that these solutions belong to
$\mathcal{D}(h_{\rho,0})$, such that it is of the form
\eqref{eq:domain} for some $\rho$ selected from the physics
of the problem.
So, we substitute \eqref{eq:sver} into \eqref{eq:domain} and
compute
$\lim_{r_{0}\to 0^{+}}r_{0}{\dot{\psi}_{\rho}}/{\psi_{\rho}}|_{r=r_{0}}$
by using the asymptotic representation for $K_{\nu}(z)$
in the limit $z\rightarrow 0$, which is given by
\begin{equation}
  K_{\nu}(z)\sim
  \frac{\pi}{2\sin (\pi \nu)}
  \left[
    \frac{z^{-\nu}}{2^{-\nu}\Gamma(1-\nu)}-
    \frac{z^{ \nu}}{2^{ \nu}\Gamma(1+\nu)}
  \right].
  \label{eq:besselasympt}
\end{equation}
After a straightforward calculation, we have the relation
\begin{align}
  \label{eq:derfe}
  \lambda = {}
  &
    \lim_{r_{0}\to 0^{+}}r_{0}
    \frac{\dot{\psi}_{\rho,0}}{\psi_{\rho,0}}\Big|_{r=r_{0}}
  \nonumber \\
    = {}
  &
  \frac
  {|j|
    \left[
      r_{0}^{2|j|}
      \Gamma(1-|j|)
      (\kappa_{b}/2)^{|j|}+
      2^{|j|} \Gamma(1+|j|)
    \right]}
  {r_{0}^{2 |j|}\Gamma(1-|j|)
    (\kappa_{b}/2)^{|j|}-
    2^{|j|} \Gamma(1+|j|)},
\end{align}
where we used Eqs. \eqref{eq:logder} and \eqref{eq:nrs}.
Then, solving the above equation for $E_{b}$, we find the sought bound
state energy
\begin{equation}
  E_{b}=-
  \frac{2}{Mr_{0}^{2}}
  \left[
     \left(
      \frac
      {\lambda+ |j|}
      {\lambda-|j|}
    \right)
    \frac
    {\Gamma(1+|j|)}
    {\Gamma(1-|j|)}
  \right]^{1/|j|}.
\label{eq:energy_KS}
\end{equation}
Now, that we have the bound state energy obtained from BG and KS methods
we can compare their results.
Thus comparing \eqref{eq:energy_BG} with \eqref{eq:energy_KS} we have
the following relation
\begin{equation}
  \label{eq:lambda_r0}
  \nu = -\frac{1}{r_0^{2|j|}}\left(\frac{\lambda+|j|}{\lambda-|j|}\right).
\end{equation}
So, we have obtained a relation between the self-adjoint extension
parameter and physical parameters of the system.

\section{Conclusions}
\label{sec:conclusion}

In this work, we have discussed the self-adjoint extension approach to
deal with singular Hamiltonians in (2+1) dimensions.
Two different methods, both based on the self-adjoint extension approach were discussed in details.
The BG and KS methods were applied to solve the problem of a spin--1/2
charged particle with an anomalous magnetic moment in the curved space.
The presence of the spin gives rise to a point interaction, requiring
the use of the self-adjoint extension approach to solving the problem.
In the BG method, the S-matrix was determined and from its poles, one bound state energy expression was obtained.
These results were obtained by imposing a suitable  boundary condition
and depend on the self-adjoint extension parameter, which can
be identified as the inverse of the scattering length of the
Hamiltonian.
Nevertheless, from the mathematical point of view, this parameter is
arbitrary.
Then, by applying the KS method, an expression for the bound state
energy for the same system was obtained, and it is given in terms of
physical parameters of the system.
Thus comparing the results from both methods a relation between the
self-adjoint extension parameter and physical parameters was
obtained.

\section*{Acknowledgments}
This work was partially supported by the Brazilian agencies
CNPq (Grants No. 313274/2017-7, No. 434134/3028-0, 427214/2016-5 and 303774/2016-9),
Funda\c{c}\~{a}o Arauc\'{a}ria (Grant No. 09/2017),
FAPEMA (Grants No. 01852/14 and No. 01202/16).
This study was financed in part by the Coordena\c{c}\~{a}o de
Aperfei\c{c}oamento de Pessoal de N\'{\i}vel Superior - Brasil (CAPES) -
Finance Code 001.

\bibliographystyle{apsrev4-2}
\bibliography{abacone-R2.bbl}

\end{document}